\documentclass[12pt,a4paper]{article}

\usepackage{a4wide}
\usepackage{amsmath,amssymb}
\usepackage{bm}
\usepackage{amssymb}
\usepackage{hyperref}
\usepackage{epic}
\usepackage{graphicx}

\newcommand{\Hamiltonian}{{\rm \bf{H}}}
\newcommand{\Dilatation}{{\rm \bf{D}}}
\newcommand{\Shift}{{\rm \bf{U}}}
\newcommand{\SpinQ}{\mathcal{S}}
\newcommand{\SpinAux}{{\bf J}}
\newcommand{\Permutation}{\mathbb{P}}
\newcommand{\Identity}{\mathbb{I}}
\newcommand{\specfad}{u}
\newcommand{\specbaz}{z}
\newcommand{\specL}{z_-}
\newcommand{\specLbar}{z_+}
\newcommand{\specsubtraction}{z_0}
\newcommand{\twist}{\phi}
\newcommand{\Qop}{{\rm \bf{Q}}}
\newcommand{\Pop}{{\rm \bf{P}}}
\newcommand{\Nop}{{\rm \bf{N}}}

\newcommand{\Monodromy}{{\rm {\mathcal M}}}
\newcommand{\Top}{{\rm {\bf T}}}

\newcommand{\Lbaz}{{\rm L}}

 \newcommand{\indone}{1}
 \newcommand{\indtwo}{2}
\newcommand{\Lbb}{{\mathbb L}}
\newcommand{\Rbb}{{\mathbb R}}
\newcommand{\Tbb}{{\mathbb T}}
\newcommand{\Dbb}{{\mathbb D}}
\newcommand{\beq}{\begin{equation}}
\newcommand{\eeq}{\end{equation}}
\newcommand{\V}{{V}}

\newcommand{\Ccal}{{\mathcal C}}
\newcommand{\Mcal}{{\mathcal M}}

\newcommand{\Tr}{{\rm Tr \,}}
\newcommand{\Op}{{\cal O}}
\newcommand{\alg}[1]{\mathfrak{#1}}
\newcommand{\superN}{\mathcal{N}}
\newcommand{\osca}{\mathbf{a}}

\newcommand{\osch}{\mathbf{h}}

\newcommand{\sfrac}[2]{{\textstyle\frac{#1}{#2}}}
\newcommand{\half}{\sfrac{1}{2}}

\makeatletter
\def\mr@ignsp#1 {\ifx\:#1\@empty\else #1\expandafter\mr@ignsp\fi}%
\newcommand{\multiref}[1]{\begingroup
\xdef\mr@no@sparg{\expandafter\mr@ignsp#1 \: }%
\def\mr@comma{}%
\@for\mr@refs:=\mr@no@sparg\do{\mr@comma\def\mr@comma{,}\ref{\mr@refs}}%
\endgroup}
\makeatother

\newcommand{\Appref}[1]{appendix~\multiref{#1}}

\numberwithin{equation}{section}

\makeatletter
\let\old@startsection=\@startsection
\renewcommand{\@startsection}[6]{\old@startsection{#1}{#2}{#3}{#4}{#5}{#6\mathversion{bold}}}
\makeatother

\begin{document}
\thispagestyle{empty}

\begin{flushright}\footnotesize
\texttt{AEI-2010-023}\\
\texttt{HU-EP-10/24}\\
\texttt{HU-Mathematik:~2010-7}\\
\vspace{0.5cm}
\end{flushright}
\setcounter{footnote}{0}

\begin{center}
{\Large{\bf A Shortcut to the Q-Operator }}
\vspace{15mm}

{\sc Vladimir V.~Bazhanov $^a$, Tomasz {\L}ukowski $^{b,c}$, Carlo Meneghelli $^c$,
Matthias Staudacher $^{c,d}$}\\[5mm]

{\it $^a$ Department of Theoretical Physics,
     Research School of Physics and Engineering\\
     Australian National University, Canberra, ACT 0200, Australia}\\[5mm]

{\it $^b$ Institute of Physics, Jagellonian University\\
     ul.~Reymonta 4, 30-059 Krak\'ow, Poland}\\[5mm]

{\it $^c$ Max-Planck-Institut f\"ur Gravitationsphysik, Albert-Einstein-Institut\\
    Am M\"uhlenberg 1, 14476 Potsdam, Germany}\\[5mm]
    
{\it $^d$ Institut f\"ur Mathematik und Institut f\"ur Physik, Humboldt-Universit\"at zu Berlin\\
Postal address: Unter den Linden 6, 10099 Berlin, Germany
}\\[5mm]

\texttt{Vladimir.Bazhanov@anu.edu.au}\\
\texttt{tomaszlukowski@gmail.com}\\
\texttt{carlo@aei.mpg.de}\\
\texttt{matthias@aei.mpg.de}\\[18mm]

\textbf{Abstract}\\[2mm]
\end{center}

\noindent{Baxter's Q-operator is generally believed to be the most powerful tool for the exact diagonalization of integrable models. Curiously, it has hitherto not yet been properly constructed in the simplest such system, the compact spin-$\half$ Heisenberg-Bethe XXX spin chain. Here we attempt to fill this gap and show how two linearly independent operatorial solutions to Baxter's TQ equation may be constructed as commuting transfer matrices if a twist field is present. The latter are obtained by tracing over infinitely many oscillator states living in the auxiliary channel of an associated monodromy matrix. We furthermore compare and differentiate our approach to earlier articles addressing the problem of the construction of the Q-operator for the XXX chain. Finally we speculate on the importance of Q-operators for the physical interpretation of recent proposals for the Y-system of AdS/CFT.}
\newpage

\setcounter{page}{1}
\section{Motivation}
\label{sec:intro}

Recently there was much progress with integrability in planar four-dimensional gauge theories \cite{Lipatov:1997vu,Braun:1998id,Minahan:2002ve} and AdS/CFT \cite{Bena:2003wd,Beisert:2005fw}. At variance with the long-held belief that quantum integrability is confined to low-dimensional systems, the asymptotic Bethe Ansatz solution of planar $\superN=4$ gauge theory was conjectured by a combination of rigorous results and assumptions \cite{Beisert:2005fw}. Further conjectures and arcane techniques led to first proposals \cite{Arutyunov:2009zu} for the full, non-asymptotic spectrum of this gauge theory, as well as its dual superstring theory on the space ${\rm AdS}_5 \times {\rm S}^5$. The proper interpretation and even the veracity of these latter proposals remains hotly debated, and in any case there is currently no trace of the theoretical underpinnings of the employed experimental mathematics. The final equations obtained take the form of an infinite system of integro-difference equations, based on a conjectured thermodynamic Bethe Ansatz, which involve an infinite set of ``$T$-functions''. These may in turn be rewritten for a sometimes more convenient set of ``$Y$-functions''. In the large volume limit the ``$T$-functions'' are to be interpreted as transfer matrices of an infinite number of excitations, most of them corresponding to bound states. However, some of the $T$-functions involve excitations believed to be elementary (mirror-magnons), while others 
turn into what looks like the eigenvalues of some Baxter $\Qop$-operators.

It is currently totally unclear which {\it operators} $\Top$ and $\Qop$ possess all these eigenvalues $T$ and $Q$. In other words, even if the proposals \cite{Arutyunov:2009zu} are made more precise, and turn out to be correct, the question {\it ``What has been diagonalized?''} will remain unanswered. One clearly would like to construct the associated operators, and prove that their mutual commutativity is based on an underlying Yang-Baxter symmetry.

One very curious feature of AdS/CFT integrability is that the underlying integrable system is both a quantum sigma-model ``living'' on a smooth continuous two-dimensional worldsheet, and at the same time a certain long-range spin chain defined on a discrete lattice. The two pictures are related by a continuous coupling constant, and one cannot say that the continuum sigma model is obtained from a discrete spin chain by a continuum limit. The sigma model {\it is} also a spin chain.

For short-range spin chains, the construction of transfer matrix
operators $\Top$ is rather well understood, see below. However, the
general principles of construction of the Baxter ${\bf Q}$-operators, originally
introduced in his seminal paper on the 8-vertex model \cite{Baxter:1972hz}, are
less well understood, and 
appear to be less systematic, even though the subject was intensively 
studied for the past twenty years.
An important step toward a general algebraic theory of the ${\bf Q}$-operators, 
particularly relevant for this paper, was made  in 
\cite{Bazhanov:1996dr,Bazhanov:1998dq} devoted to conformal field theory
where  these operators were constructed as {\em traces of certain
monodromy matrices}, associated with infinite dimensional
representations of the $q$-deformed harmonic oscillator algebra. More
generally this method applies to any model with $U_q(\widehat{sl}(2))$
symmetry and it was further generalized for the case of some
higher-rank 
quantized algebras and super-algebras \cite{Bazhanov:2001xm, Bazhanov:2008yc, Kojima08}.
However despite all these considerations of rather complicated models
with ``$q$-deformed'' symmetries, it appears that there is 
still no completely explicit construction
in the literature for the ${\bf Q}$-operator
 of the compact XXX chain --- the first spin chain ever
solved by Bethe Ansatz \cite{Bethe}.
We therefore decided to fill this gap, as a very first
necessary step to vigorously address the much more involved case of AdS/CFT
integrability. Excitingly, we will find that we need infinite
dimensional representations to carry out this construction. These are
therefore needed to properly understand the integrable structure of
spin chains with finite dimensional quantum space, a feature which
then puts spin chains on a par with integrable sigma models. We
believe that this makes the appearance of spin chains in the strongly
coupled limit of the AdS/CFT sigma model much less surprising.

\section{Brief Review of the Spin-\texorpdfstring{$\half$}{} XXX Heisenberg Chain}
\label{sec:review}

The one-dimensional Heisenberg spin chain Hamiltonian reads
\begin{equation}\label{hamiltonian}
\Hamiltonian=4\, \sum_{l=1}^{L}\left( \frac{1}{4}-\vec \SpinQ_{l} \cdot \vec \SpinQ_{l+1}\right)
\qquad {\rm with} \qquad 
\vec \SpinQ =\frac{1}{2}\, \vec \sigma\, ,
\end{equation}
where $\vec \sigma$ are the three Pauli matrices, i.e.~$\vec \SpinQ$ is the spin-$\half$ representation of $\alg{su}(2)$. This Hamiltonian acts on the $L$-fold tensor product 
\begin{equation}\label{quantumspace}
{\mathcal V}=\underbrace{
{\mathbb C}^2\otimes {\mathbb C}^2\otimes \cdots \otimes  
{\mathbb C}^2}_{L-\mbox{\scriptsize{times}}}
\ ,
\end{equation}
which will be called {\it quantum space} throughout the paper. The spin operator $\vec \SpinQ_l$ acts on the $l$-th component of the quantum space, and it is clear from \eqref{hamiltonian} that we need to specify the meaning of $\vec \SpinQ_{L+1}$. Periodic boundary conditions are imposed by defining 
\begin{equation}\label{periodic}
\vec \SpinQ_{L+1}:=\vec \SpinQ_1
\, .
\end{equation}
This Hamiltonian may be rewritten as 
$\Hamiltonian=2\, \sum_{l=1}^{L}\left(\Identity_{l,l+1}-\Permutation_{l,l+1}\right)$ where $\Identity_{l,l+1}$ and $\Permutation_{l,l+1}$ are the identity and the spin permutation operators on adjacent sites $(l,l+1)$ of the chain of length $L$, respectively. It appears in $\superN=4$ Yang-Mills theory in the scalar field subsector, where $\vec \SpinQ \in \alg{su}(2) \subset \alg{su}(4) \subset \alg{psu}(2,2|4)$, as the one-loop approximation of the conformal dilatation generator $\Dilatation \in \alg{su}(2,2) \subset \alg{psu}(2,2|4)$  
\begin{equation}\label{dilatation}
\Dilatation=L+g^2\, \Hamiltonian+\Op(g^4)\, .
\end{equation}
Here $g^2$ is related to the `t Hooft coupling constant $\lambda$ by $g^2=\frac{\lambda}{16 \pi^2}$. Note that the Hamiltonian \eqref{hamiltonian} with boundary conditions \eqref{periodic} is rotationally invariant, i.e.~$[ \Hamiltonian, \vec \SpinQ] =0$. 

It is well known that Hans Bethe discovered in 1931 a system of algebraic equations which yield the exact spectrum of $\Hamiltonian$. This was obtained after making an Ansatz for the wave function now carrying his name \cite{Bethe}.  This so-called coordinate Bethe Ansatz interprets the state 
$\uparrow \ldots \uparrow$
with energy eigenvalue $E=0$ as the vacuum. Each up-spin $\uparrow$ is an unoccupied lattice site, and each down-spin 
$\downarrow$ is interpreted as a lattice particle, termed magnon, carrying lattice momentum $p$. After introducing a rapidity $\specfad$ for each magnon, where 
\begin{equation}\label{raptomom}
e^{i p}=\frac{\specfad+\frac{i}{2}}{\specfad-\frac{i}{2}}\ \ \  \Longleftrightarrow \ \ \ \specfad=\frac{1}{2}\cot \frac{p}{2}\, ,
\end{equation}
Bethe's solution for the eigenvalues $E$ of $\Hamiltonian$ in the (conserved) sector of $M$ magnons reads
\begin{equation}\label{energy}
E=2\,\sum_{k=1}^{M}\frac{1}{\specfad_{k}^{2}+\frac{1}{4}}\, ,
\end{equation}
where the $M$ Bethe roots $\specfad_k$ have to satisfy the Bethe equations
\begin{equation}\label{betheeqsnotwist}
\Big( \frac{\specfad_k+\frac{i}{2}}{\specfad_k-\frac{i}{2}}\Big) ^L =\prod_{\stackrel{j=1}{j\neq k}}^{M} \frac{\specfad_k-\specfad_j+i}{\specfad_k-\specfad_j-i}\, .
\end{equation}
The eigenvalue U of the lattice translation operator $\Shift$, which shifts a given spin configuration by one lattice site, is given by
\begin{equation}\label{shift}
U=
\prod_{k=1}^M \frac{\specfad_k+\frac{i}{2}}{\specfad_k-\frac{i}{2}}
\, .
\end{equation}

There has been a longstanding controversy, starting from a discussion
in Bethe's original paper, as to whether the obtained spectrum is
complete, which still attracts much attention; here we give only a small
sample of references \cite{Kirillov:1994cp,Baxter:2001sx,Fabricius:2000yx,mukhin}. There are (at least) three distinct subtleties. The first is, that the map \eqref{raptomom} leads to infinite rapidities $\specfad=\infty$ for zero-momentum magnons where $p=0$. Of course this is consistent with both the expression for the energy \eqref{energy} and the Bethe equations \eqref{betheeqsnotwist}, but complicates the proper counting of states. The effect may be traced to the $\alg{su}(2)$ invariance of $\Hamiltonian$: The highest weight state of each energy multiplet corresponds to a solution with only finite rapidities. The descendents of this state are obtained by applying the global $\alg{su}(2)$ lowering operator. Each application places a further $p=0$ magnon into the chain, corresponding to a completely symmetrized insertion. A second subtlety is that it is a priori not allowed to increase $M$ beyond half-filling, i.e.~such that $M>L/2$ (all states must then be descendents). Spurious states with finite rapidities are nevertheless found. In fact, one ``experimentally'' finds that solving the equations 
\begin{equation}\label{dualbetheeqsnotwist}
\Big( \frac{\tilde{\specfad}_k+\frac{i}{2}}{\tilde{\specfad}_k-\frac{i}{2}}\Big) ^L 
=
\prod_{\stackrel{j=1}{j\neq k}}^{L-M+1} \frac{\tilde{\specfad}_k-\tilde{\specfad}_j+i}{\tilde{\specfad}_k-\tilde{\specfad}_j-i}
\end{equation}
for $L-M+1$ roots $\tilde u_k$ and plugging the solution  into
\begin{equation}\label{dualenergynotwist}
E=2\sum_{k=1}^{L-M+1}\frac{1}{\tilde{\specfad}_{k}^{2}+\frac{1}{4}}
\end{equation}
yields energies identical to the ones of highest weight states with magnon number $M$. However, the root distribution $\tilde \specfad_k=\tilde \specfad_k(\alpha)$ depends on one arbitrary complex parameter $\alpha$, while the energy does not. These ``beyond the equation solutions'' were discussed in detail in \cite{Pronko:1998xa}.
The third subtlety, already noted in \cite{Bethe}, is that some momenta appear in pairs of the form $p=p_0  \pm i\, \infty$, which leads via \eqref{raptomom} to Bethe roots at $\specfad=\pm \frac{i}{2}$. The physical picture here is that two magnons form an infinitely tight bound state\footnote{This happens first for one of the two singlet states of the $L=4$ spin chain.}, they are ``stuck together''. However, they of course contribute only a finite amount of energy, so \eqref{energy} has to be interpreted with great care.

It is also quite well known that all these subtleties may be resolved by replacing 
$\Hamiltonian$ by a ``twisted'' Hamiltonian
$\Hamiltonian_\twist$, where $\phi$ can be interpreted as an (imaginary) ``horizontal field'' in condensed matter parlance,
or alternatively as a magnetic flux passing through the chain looped into a circle
\begin{equation}\label{hamiltonian2}
\Hamiltonian_\twist=4\, \sum_{l=1}^{L}\big[ \frac{1}{4}-\SpinQ_{l}^{3}\,\SpinQ_{l+1}^{3}-\frac{1}{2}\,e^{i\frac{\twist}{L}}\,\SpinQ_{l}^{+}\,\SpinQ_{l+1}^{-}-\frac{1}{2}\,e^{-i\frac{\twist}{L}}\,\SpinQ_{l}^{-}\,\SpinQ_{l+1}^{+}\big]\, ,
\end{equation}
with $\SpinQ_l^{\pm}=\SpinQ_l^1\pm i\, \SpinQ_l^2$.
The Bethe Ansatz still works with minor modifications. The formula for the energy \eqref{energy} and the relations \eqref{raptomom} remain unaffected, but the Bethe equations \eqref{betheeqsnotwist} are modified to
\begin{equation}\label{betheeqs}
\Big( \frac{\specfad_k+\frac{i}{2}}{\specfad_k-\frac{i}{2}}\Big) ^L e^{i\twist}=\prod_{\stackrel{j=1}{j\neq k}}^{M} \frac{\specfad_k-\specfad_j+i}{\specfad_k-\specfad_j-i}\, .
\end{equation}
This tiny modification resolves all the difficulties we just discussed (for generic values of the twist). Firstly one now finds {\it all} $2^L$ states of the length $L$ spin chain, such that all rapidities $\specfad_k$ are finite. Secondly, there is no ``beyond the equator problem" anymore and $M$ can range over
$M\in\{0,1,\ldots,L-1,L\}$. In fact one can now rigorously derive the second Bethe Ansatz, which treats the up-spins as ``dual'' magnons in the vacuum of the down-spins.  It reads 
\begin{equation}\label{dualbetheeqs}
\Big( \frac{\tilde{\specfad}_k+\frac{i}{2}}{\tilde{\specfad}_k-\frac{i}{2}}\Big) ^L e^{-i\phi}=\prod_{\stackrel{j=1}{j\neq k}}^{L-M} \frac{\tilde{\specfad}_k-\tilde{\specfad}_j+i}{\tilde{\specfad}_k-\tilde{\specfad}_j-i}\, ,
\end{equation}
\begin{equation}\label{dualenergy}
E=2\sum_{k=1}^{L-M}\frac{1}{\tilde{u}_{k}^{2}+\frac{1}{4}}\, ,
\end{equation}
and in contradistinction to \eqref{dualbetheeqsnotwist},\eqref{dualenergynotwist} there are only $L-M$ dual magnons, as expected. In fact, the set of equations \eqref{betheeqs} and \eqref{dualbetheeqs} are completely equivalent, each set yields the full spectrum of all $2^L$ states.
Finally there are no more degenerate roots 
at $\specfad=\pm \frac{i}{2}$ with a priori indefinite energy \eqref{energy} or \eqref{dualenergy}.
The price to pay is that $\alg{su}(2)$ invariance of the spectrum is broken: All multiplets split up\footnote{A spin $s$ multiplet where the magnetic quantum numbers are $m=-(2s+1), \ldots, (2 s+1)$ splits up such that there still is a degeneracy between any two states whose $m$ differs by a sign flip.}. However, we can think of $\twist$ as a small regulator which may always be removed where physically sensible.  Note that such twists are also natural from the point of view of the AdS/CFT correspondence. They appear in the scalar sector of the integrable one-loop dilatation operator of the  $\beta$-deformed twisted ${\cal N}=4$ gauge theory \cite{Beisert:2005if}.

The magnetic flux $\phi$ may be distributed in many possible ways. For instance, as concerns the energy spectrum of the spin chain, it is equivalent to use instead of \eqref{hamiltonian2} a linearly transformed Hamiltonian, 
\beq
\widetilde{\Hamiltonian}_\twist=
\Ccal\big(\twist/L\big)\,{\Hamiltonian}_\twist \;
\Ccal\big({\twist}/{L}\big)^{-1},\qquad 
\Ccal(\alpha)=e^{i\,L\alpha\,\SpinQ^3_L}\otimes
e^{i\,(L-1)\,\alpha\,\SpinQ^3_{L-1}} \otimes\cdots\otimes
e^{i\,\alpha\,\SpinQ_1^3}\ , \label{Htilde}
\eeq
which is given by the original formula \eqref{hamiltonian}, but with
``twisted'' boundary conditions
\begin{equation}\label{twisted}
\SpinQ^3_{L+1}:= \SpinQ^3_1\, ,
\qquad \qquad
\SpinQ^{\pm}_{L+1}:=e^{\mp i \phi}\, \SpinQ^{\pm}_1
\, .
\end{equation}

There is an interesting way to reformulate the Bethe equations \eqref{betheeqs} and \eqref{dualbetheeqs} with the help of Baxter polynomials defined for each eigenstate by
\begin{equation}\label{baxterpoly}
A_-(\specfad):=\prod_{k=1}^M (\specfad-\specfad_k)\, ,
\qquad \qquad
A_+(\specfad):=\prod_{k=1}^{L-M} (\specfad-\tilde \specfad_k)
\, .
\end{equation}
Then
\begin{equation}\label{betheeqsfromA}
\Big( \frac{\specfad+\frac{i}{2}}{\specfad-\frac{i}{2}}\Big) ^L e^{\pm i\twist}=
-\frac{A_{\mp}(\specfad+i)}{A_{\mp}(\specfad-i)}
\qquad {\rm if} \qquad
\specfad=\specfad_k
\quad {\rm or} \quad
\specfad=\tilde \specfad_k\, ,
\quad {\rm respectively.}
\end{equation}
On infers that the following equations must hold for each eigenstate and for all $\specfad \in \mathbb{C}$:
\begin{equation}\label{BaxtereqA}
T(\specfad)\,A_{\mp}(\specfad)
=
e^{\pm i\frac{\twist}{2}}\, \left(\specfad+\frac{i}{2}\right)^L\,A_{\mp}(\specfad-i) +
e^{\mp i\frac{\twist}{2}}\, \left(\specfad-\frac{i}{2}\right)^L\,A_{\mp}(\specfad+i) 
\, .
\end{equation}
The reason is that the r.h.s.~is a polynomial with roots at $\specfad=\specfad_k$ or $\specfad=\tilde \specfad_k$, respectively, so it must be proportional to $A_{\mp}(\specfad)$. We also see that the latter function must be multiplied by another polynomial $T_{\pm}(\specfad)$ of degree $L$. One experimentally finds that this further polynomial does {\it not} depend on $\pm$, i.e.~$T_{\pm}(\specfad)=T(\specfad)$, a fact which will be proven in the next chapter. So we have from \eqref{BaxtereqA} and \eqref{baxterpoly}
\begin{equation}\label{Tpoly}
T(\specfad)=2\,\cos \frac{\twist}{2}\,\prod_{k=1}^L (\specfad-w_k)
\, ,
\end{equation}
where the $w_k$ are the remaining $L$ roots of the r.h.s.~of \eqref{BaxtereqA}.

Equation \eqref{BaxtereqA} can be regarded as a second order difference equation for an unknown function $A(\specfad)$, which has two linearly independent solutions $A_\pm(\specfad)$. To see this more clearly, it is convenient to define the Baxter functions
\begin{equation}\label{baxterfunction}
Q_{\pm}(\specfad):=e^{\pm u\frac{\twist}{2}}\,A_{\pm}(\specfad)
\, ,
\end{equation}
such that $Q_{\pm}(\specfad)$ are indeed two linearly independent
solutions\footnote{The most general {\it formal} solution of
  \eqref{BaxtereqQ} is then a linear superposition of the form
  $Q(\specfad)=c_+(\specfad)\,Q_+(\specfad)+c_-(\specfad)\,Q_-(\specfad)$.
  However, unlike the theory of second order differential (as opposed
  to difference) equations is that the ``constants''
  $c_{\pm}(\specfad)$ could a priori be any functions of $\specfad$
  with period $i$.}  of the difference equation
%
%
%
\begin{equation}\label{BaxtereqQ}
T(\specfad)\,Q(\specfad)
=
\left(\specfad+\frac{i}{2}\right)^L\,Q(\specfad-i) +
\left(\specfad-\frac{i}{2}\right)^L\,Q(\specfad+i)
\quad {\rm with} \quad
Q(\specfad)=Q_{\pm}(\specfad)
\, .
\end{equation}
This is Baxter's famous $TQ$ equation for the twisted Heisenberg
magnet. As we can see, the twist has actually disappeared from the
equation, and is entirely encoded in the analytic Ansatz
\eqref{baxterfunction},\eqref{baxterpoly},\eqref{Tpoly} for the
solution. Note that the Baxter functions at nonzero twist $\twist$ are
{\it not} polynomials.

Baxter derived this equation, which holds for all eigenvalues of the
commuting ${\bf T}$- and ${\bf Q}$-matrices on the operatorial level,
in his original solution on the ``zero-field'' 8-vertex model
\cite{Baxter:1972hz}, which also contains the solution of the
``zero-field'' XYZ spin chain \cite{Baxter:1972wg}.  Although it seems to
be possible to take a limit of his results and apply them to the untwisted XXX model, this would then only apply to the ``zero-field''  case
$\phi=0$.

One purpose of this article is to consider the XXX chain with an
arbitrary non-zero field $\phi\not=0$ and provide an independent
construction of the {\it operators} $\Top({\specfad})$ and
$\Qop_{\pm}(\specfad)$ satisfying \eqref{BaxtereqQ} such that all
their eigenvalues are of the form spelled out in
\eqref{baxterfunction}, \eqref{baxterpoly} and \eqref{Tpoly}.  Our
construction of the ${\bf Q}$-operators is conceptually very similar
to that of \cite{Bazhanov:1996dr,Bazhanov:1998dq} in $c<1$ conformal field theory, based on
the Yang-Baxter equation with $U_q(\widehat{sl}(2))$ symmetry.
However, here we provide a self-contained and separate consideration of
the XXX model, rather than attempting to take the $q^2\to1$ limit of the
relevant results of \cite{Bazhanov:1996dr,Bazhanov:1998dq}.

In the case of the operator $\Top(\specfad)$ Baxter's construction
actually immediately applies. His methodology was subsequently
developed and systematized by the inverse scattering methodology of
the Leningrad school of mathematical physics. Here we will just
briefly summarize the construction, referring for all further details
to the authoritative presentation \cite{Faddeev:1996iy}. One
constructs this operator, termed {\it transfer matrix}
$\Top(\specfad)$, as the trace over a certain {\it monodromy matrix}
$\Monodromy(\specfad)$.  The latter is in turn built from a
``generating object'', the local quantum {\it Lax operator}
\begin{equation}\label{laxfaddeev}
\mathcal{L}_l(\specfad)=
\left( \begin{array}{cc}
\specfad+ i\,\SpinQ_l^{3} &  ~~i\,\SpinQ_l^- \\
i\,\SpinQ_l^+ & ~~\specfad-i\,\SpinQ_l^{3}  \end{array} \right),
\end{equation}
which is a $2 \times 2$ matrix in some auxiliary space $\mathbb{C}^2$ with site-$l$ spin operator (as in \eqref{hamiltonian}) valued matrix elements. The monodromy matrix is then built as\footnote{Here $\cdot$ denotes $2 \times 2$ matrix multiplication in the two-dimensional auxiliary space. The entries of this $2 \times 2$ matrix act on \eqref{quantumspace}.}
\begin{equation}\label{monodromy}
\Monodromy(\specfad)=
\left( \begin{array}{cc}
e^{i\,\frac{\twist}{2}} & 0 \\
0  & e^{-i\,\frac{\twist}{2}}  \end{array} \right)\cdot
\mathcal{L}_L(\specfad)\cdot\mathcal{L}_{L-1}(\specfad)\cdot \ldots \cdot \mathcal{L}_2(\specfad)\cdot\mathcal{L}_1(\specfad)
\, .
\end{equation}
In view of \eqref{quantumspace} it is an operator acting on the tensor product of the auxiliary space and the quantum space. Finally, by taking the trace over the two-dimensional auxiliary space, 
\begin{equation}\label{transfertrace}
\Top(\specfad)=\Tr \Monodromy(\specfad)
\, ,
\end{equation}
one constructs the transfer matrix as an operator on the quantum space \eqref{quantumspace}. It is easy to show, see \cite{Faddeev:1996iy}, that at the special point $\specfad=\frac{i}{2}$ the transfer matrix becomes proportional to the lattice shift operator (cf.~discussion around \eqref{shift}), multiplied by a diagonal matrix, i.e.~one has 
%
%
%
\begin{equation}\label{shiftfromT}
{\bf U}=
i^{-L}\,\Top(\tfrac{i}{2})\,
e^{-i\,\twist\,S^3_{L}}\,
\, .
\end{equation}
The Hamiltonian \eqref{Htilde} 
is then obtained from the expansion of the transfer matrix in the 
vicinity of the point $\specfad=\frac{i}{2}$\ , 
\begin{equation}\label{HfromT}
\widetilde{\Hamiltonian}_\twist
=2\,L-2\,i\,\frac{d}{d u} \log \Top(\specfad) \Big|_{\specfad=\frac{i}{2}}
\, .
\end{equation}
Finally, because of the underlying Yang-Baxter symmetry, to be
discussed below, one can show that transfer matrices with different
values of spectral parameters form a commuting family 
\begin{equation}\label{Tcommutes}
\left[  \Top(\specfad),  \Top(\specfad') \right]=0
\, ,
\end{equation}
which also contains the Hamiltonian \eqref{Htilde}, obtained from
\eqref{hamiltonian} by twisting the boundary conditions
\eqref{twisted}. 
However, this fact by itself does not directly lead to the solution of the model. One procedure is to apply the algebraic Bethe Ansatz as explained in \cite{Faddeev:1996iy}. Another, to be developed below, is to also construct the operators $\Qop_{\pm}(\specfad)$ as traces over some monodromy matrices, and to derive the operator version of \eqref{BaxtereqQ}
\begin{equation}\label{OpBaxtereqQ}
\Top(\specfad)\,\Qop_{\pm}(\specfad)
=
\left(\specfad+\sfrac{i}{2}\right)^L\,\Qop_{\pm}(\specfad-i) +
\left(\specfad-\sfrac{i}{2}\right)^L\,\Qop_{\pm}(\specfad+i)
\, ,
\end{equation}
along with a proof that the analyticity of the eigenvalues of the therein appearing operators is given by \eqref{baxterfunction},\eqref{baxterpoly},\eqref{Tpoly}. Then no Bethe Ansatz is required, and the twisted Bethe equations \eqref{betheeqs} or \eqref{dualbetheeqs} immediately follow (for more details on this logic, please see section \ref{sec:bethe} below).

A very interesting issue is the $\twist \rightarrow 0$ limit of the operators $\Qop_{\pm}(\specfad)$ appearing in \eqref{OpBaxtereqQ}. In this limit, the broken $\alg{su}(2)$ invariance of the spin chain is recovered. Then the majority of the eigenvalues of $\Qop_{\pm}(\specfad)$ turn into  descendents of $\alg{su}(2)$ highest weight states. From our discussion following the untwisted Bethe equations \eqref{betheeqsnotwist} these will have Bethe roots at $\specfad=\infty$, so the operators $\Qop_{\pm}(\specfad)$  are expected to diverge, in stark contrast to the eigenvalues of $\Top(\specfad)$, which are perfectly finite and smooth in the zero-twist limit. On the other hand, we can of course retrace the steps leading to the existence of the operator equations \eqref{OpBaxtereqQ} by making, in analogy with \eqref{baxterpoly}, a polynomial Ansatz for linearly independent Baxter functions\footnote{Note that the exponential factors in \eqref{baxterfunction} disappear at $\twist=0$. We have used the notation $\sim$ since the proper normalization of these functions will turn out to be quite subtle.}
\begin{equation}\label{baxterpolyPQ}
Q(\specfad) \sim \prod_{k=1}^M (\specfad-\specfad_k),
\qquad \qquad
P(\specfad)\sim\prod_{k=1}^{L-M+1} (\specfad-\tilde \specfad_k).
\end{equation}
This empirical way had already been explored in \cite{Pronko:1998xa}, but it was erroneously (as we shall see) stated
that $P(\specfad)$ and $Q(\specfad)$ are completely unrelated to the eigenvalues of $\Qop_\pm$. 
By using the untwisted Bethe \eqref{betheeqsnotwist} and dual Bethe \eqref{dualbetheeqsnotwist} equations, the latter being the reason for the power $L-M+1$ in the polynomial $P(\specfad)$, we 
conclude that there must be {\it finite operators} $\Qop(\specfad), \Pop(\specfad)$ which should satisfy the {\it same} Baxter equation \eqref{OpBaxtereqQ} as $\Qop_{\pm}(\specfad)$, such that however (1) their eigenvalue spectrum is $\alg{su}(2)$ invariant, and (2) their eigenvalues are indeed of the form in \eqref{baxterpolyPQ}. It is clear that finding $\Qop(\specfad), \Pop(\specfad)$ from $\Qop_{\pm}(\specfad)$ must indeed be quite non-trivial, and must involve some kind or ``renormalization" of these divergent operators. It is furthermore a priori quite mysterious how the extra $(L-M+1)$-th root appears in $P(\specfad)$ in \eqref{baxterpolyPQ} as compared to the $L-M$ roots of $A_+(\specfad)$ in \eqref{baxterpoly}. These puzzles will be resolved in chapter \ref{sec:notwist}.
In particular, we shall find the resolution to be intimately connected to the exponential factors in \eqref{baxterfunction}.

Our article is organized as follows. In the ensuing chapter \ref{sec:construction} we will construct the operators $\Qop_{\pm}$ as the trace over an appropriate monodromy matrix. This will require the introduction of two copies of infinite oscillator Fock spaces, despite the fact that we are dealing with a finite dimensional spin chain carrying finite dimensional representations of $\alg{su}(2)$. In the next chapter \ref{sec:notwist} we will study the very subtle $\twist \rightarrow 0$ limit of our construction, leading to a one-parameter family of linearly independent operators $\Qop(\specfad), \Pop(\specfad)$ by ``renormalizing'' the previously obtained operators $\Qop_{\pm}$. We will also perform some numerical study of their spectrum for a few cases, illustrating an interesting pattern between their respective eigenvalue root distributions. In the following chapter \ref{sec:earlierwork} we briefly discuss how our construction and result differ from a large earlier, complementary literature on the subject. We end in \ref{sec:openproblems} with a brief list of the many open problems and potential applications related to our result.

Before proceeding with the construction of the operators $\Qop_{\pm}$, let us however change notation by ``Wick-rotating'' the spectral parameter $\specfad$ to
\begin{equation}\label{spectralparameters}
\specbaz:=-i\,\specfad\, .
\end{equation}
It is true that the $\specfad$-convention of this review chapter \ref{sec:review} is  the most widely accepted notation in much of the Bethe Ansatz literature, and nearly all of the literature on AdS/CFT integrability. But the $z$-convention \eqref{spectralparameters} used for the rest of this paper (with the exception of the numerical work presented in section \ref{sec:numerics} and appendix \ref{app:numerics}, as well as the examples of small chain lengths in appendix \ref{app:smalllengths}) prevents all further derivations and manipulations of functional equations from being cluttered by factors of $i$.
This will then turn \eqref{OpBaxtereqQ} into\footnote{For simplicity, and with slight abuse of notation, we will not use new symbols for the various operators. E.g. the transfer matrix in \eqref{OpBaxtereqQ} is $i^L$ times the transfer matrix in \eqref{OpBaxtereqQalt}. It should be straightforward to return to the notation with the spectral parameter $\specfad$ whenever needed by using \eqref{spectralparameters}. }
\begin{equation}\label{OpBaxtereqQalt}
\Top(\specbaz)\,\Qop_{\pm}(\specbaz)
=
\left(\specbaz+\half\right)^L\,\Qop_{\pm}(\specbaz-1) +
\left(\specbaz-\half\right)^L\,\Qop_{\pm}(\specbaz+1)
\, .
\end{equation}
%

\section{Construction of \texorpdfstring{$\Qop_\pm$}{} as Transfer Matrices}
\label{sec:construction}
\subsection{Yang-Baxter equation and commuting transfer matrices}
As explained above the Hamiltonian \eqref{hamiltonian}, with twisted
boundary conditions \eqref{twisted}, is generated by the spectral parameter-dependent transfer matrices \eqref{transfertrace}, which form a commuting family, cf.~\eqref{Tcommutes}. Here we want to construct further
important transfer matrices which nevertheless belong to the same family. 
To this end one needs to study the possible solutions of the Yang-Baxter equation 
\begin{equation}
\label{YB0}
\Rbb(x-y)\, \big(\Lbb_\V(x) \otimes 1\big)\,
\big(1\otimes \Lbb_\V(y)\big)= 
\big(1\otimes \Lbb_\V(y)\big)\,\big(\Lbb_\V(x)\otimes 1\big)\, \Rbb(x-y),
\end{equation}
where  $\Rbb(\specbaz)$ is the rational $4 \times 4$ $R$-matrix, 
\begin{equation}
{\Rbb}(\specbaz):\qquad 
{\mathbb C}^2 \otimes {\mathbb C}^2 \to {\mathbb C}^2 \otimes
{\mathbb C}^2, \qquad 
{\Rbb}(\specbaz)
=\specbaz + \Permutation\, ,\label{Ryang}
\end{equation}
and the $L$-operator\  ${\Lbb}_V(z)$ is a $2\times 2$ 
matrix, acting in the quantum space of a single spin-$\half$, 
\beq
{\Lbb}_V(z)=
\left( \begin{array}{cc}
\Lbb_{11}(z)&\quad\Lbb_{12}(z)\\
\Lbb_{21}(z)&\quad \Lbb_{22}(z)\end{array}\right)\ .
\eeq
Its matrix elements are operator-valued functions of the 
variable $z$, acting in an auxiliary vector space $V$. 
The $R$-matrix acts in a direct product of two-dimensional spaces 
${\mathbb C}^2 \otimes {\mathbb C}^2$ and the operator ${\mathbb P}$  
permutes the factors in this product. Note that the $R$-matrix \eqref{Ryang} 
is GL$(2)$-invariant
\beq
{\Rbb}(\specbaz)=(G\otimes G)\, {\Rbb}(\specbaz)\, (G\otimes G)^{-1},
\qquad G\in {\rm GL}(2),\label{gl2inv}
\eeq
where $G$ is any non-degenerate $2\times2$ matrix.

The solutions of \eqref{YB0} which we will use in this paper
are rather simple. The first one is a well-known 
generalization of \eqref{laxfaddeev},
\begin{equation}
\label{sl2invlax}
{L}(\specbaz)=\left( \begin{array}{cc}
\specbaz+ \SpinAux^{3} &  \SpinAux^- \\
\SpinAux^+ & \specbaz-\SpinAux^{3}  \end{array}
\right)=z\,\mathbb{I}+2\sum_{k=1}^3 \SpinQ^k \,\SpinAux^k ,
\end{equation}
where $\SpinAux^{\pm}=\SpinAux^1\pm i\SpinAux^2$ and $\SpinAux^{3}$ 
are the generators of the $\alg{sl}(2)$ algebra
\begin{equation}
\label{sl2commrel}
\left[ \SpinAux^3, \SpinAux^\pm \right]= 
\pm  \SpinAux^\pm\, , \qquad \qquad  
\left[ \SpinAux^+, \SpinAux^- \right]= 2\,  \SpinAux^3 \, .
\end{equation}
In the second form of $L(z)$ in \eqref{sl2invlax} we
used the $2\times2$ spin operators $\SpinQ^k$ appearing in \eqref{hamiltonian}.  
Note that for the $L$-operator \eqref{sl2invlax} equation \eqref{YB0} holds on the algebraic level in virtue of
the commutation relations \eqref{sl2commrel}.  
To obtain a specific solution one
needs to choose a particular representation of these commutation relations.
For further reference 
define the highest weight representions of $\alg{sl}(2)$, with highest weight vector $v_0$, defined by the conditions 
\beq
\SpinAux^+ v_0=0, \qquad \SpinAux^3 v_0=j\,v_0\,,\label{h-weight}
\eeq
where $j$ is the spin. The $(2j+1)$-dimensional representations 
with integer or half-integer  spin, i.e.~$2j\in {\mathbb Z}_{\ge0}$,\  will be denoted by 
$\pi_j$, while infinite-dimensional representations with arbitrary complex spin $2j\in
{\mathbb C}$ with be denoted as $\pi_j^+$. 

For each solution of \eqref{YB0} one can define a transfer matrix 
\beq
{\Tbb}_\V(z)=
{\Tr}_\V\Big({\Dbb}\,\underbrace{
{\Lbb}_\V(z)\otimes {\Lbb}_\V(z)\otimes \cdots \otimes  
{\Lbb}_\V}_{L-\mbox{\scriptsize{times}}} \Big)\ ,\label{tmat}
\eeq
where the tensor product is taken with respect to the quantum spaces 
${\mathbb C}^2$, while the operator product and the trace is taken
with respect to the auxiliary space $\V$. The boundary twist operator
$\Dbb$, specially suited for our purposes, is defined by 
\beq
[\,\Dbb,\,\Lbb_{11}(z)\,]=[\,\Dbb,\,\Lbb_{22}(z)\,]=0,\quad
\Dbb \, \Lbb_{12}(z)=e^{i\phi}\,\Lbb_{12}(z)\,\Dbb,\quad
\Dbb \, \Lbb_{21}(z)=e^{-i\phi}\,\Lbb_{21}(z)\,\Dbb\label{twistop}\, ,  
\eeq
where $\Lbb_{ij}(z)$, $i,j=1,2$, denotes matrix elements of $\Lbb_\V(z)$.
Note that $\Dbb$ is an operator in the auxiliary space; it acts
trivially in quantum space\footnote{Actually, all
  considerations presented in this paper are valid for a more general
  case, when the operator $\Dbb$ acts non-trivially in the quantum space.
The only additional condition is that it should commute with the diagonal
elements ${\mathcal M}_{jj}(z)$, $j=1,2$ of the monodromy matrix
\eqref{genmonodromy}. For example the twist could be of the form 
$\phi=\alpha+\beta\SpinQ^3_{tot}$, where $\alpha$ and $\beta$ are
$c$-numbers. This type of twist originally arose in
\cite{Bazhanov:1994ft} in  considerations of commuting
${\bf T}$-operators in CFT.}.
For the solution \eqref{sl2invlax} one easily finds
\beq
\Dbb=e^{i\phi\,\SpinAux^3}\ .\label{sl2twist}
\eeq
Substituting the last expression together with \eqref{sl2invlax} 
into the definition \eqref{tmat} and taking 
the trace over the standard $(2j+1)$-dimensional representations $\pi_j$,
one obtains an infinite set of transfer matrices
\beq\label{Tj}
{\mathbf T}_j(z)=\Tr_{\pi_j}\,\big(\Mcal(z)\big),
\eeq
built from the monodromy matrices
\beq\label{genmonodromy}
\Mcal(z)=e^{i\phi \SpinAux^3}\underbrace{
{L}(z)\otimes L(z)\otimes \cdots \otimes  
L(z)}_{L-\mbox{\scriptsize{times}}}\,,\qquad 2j\in{\mathbb Z}_{\ge0},
\eeq
and labelled by the value of the spin $j=0,\half,1,\frac{3}{2},\ldots,\infty$.
Note in particular that\footnote{%
As opposed to
\eqref{laxfaddeev} and \eqref{monodromy}, the  
$2 \times 2$ matrices $L(z)$ in \eqref{genmonodromy} act on the
corresponding copy of $\mathbb{C}^2$ in the quantum space 
\eqref{quantumspace}.}
\beq
{\mathbf T}(z)\equiv{\mathbf T}_{\frac{1}{2}}(z)
\eeq
coincides, up to a trivial rescaling of the argument $z=-iu$,
with the transfer matrix ${\mathbf T}(u)$
defined in \eqref{transfertrace}. The transfer matrices \eqref{Tj} depend 
on the spectral variable $z$ and (implicitly) on the twist parameter $\phi$.

Standard arguments based on the Yang-Baxter equation  
immediately imply that the operators ${\mathbf T_j(z)}$ belong to a commuting family, since one derives from \eqref{YB0} in generalization of \eqref{Tcommutes} that\footnote{%
One can prove that the more general relations $[{\mathbf T_j(z)},{\mathbf T_{j'}(z')}]=0$ with $2j, 2 j'\in{\mathbb Z}_{\ge0}$ also hold.
}
\beq
[{\mathbf T(z)},{\mathbf T_j(z')}]=0,\qquad 
2j\in{\mathbb Z}_{\ge0}\ .\label{Tcomm}
\eeq
The presence of the boundary twist \eqref{sl2twist} in the
definition \eqref{Tj} does not affect the commutativity thanks to the
properties \eqref{twistop} and \eqref{gl2inv}.
 
In similarity to \eqref{Tj} we may define transfer matrices
\beq
{\mathbf T}^+_j(z)=\Tr_{\pi^+_j}\,\big(\Mcal(z)\big),\qquad 
2j\in{\mathbb C},
\label{Tjplus} \eeq
where $\Mcal(z)$ is the same as in \eqref{genmonodromy}, but now the trace is taken
over an infinite-dimensional representation\footnote{%
In general this representation is not unitary, but this is not relevant to our present construction.
}
$\pi_j^+$ with an arbitrary, possibly complex spin $j$. This representation is spanned by the vectors
$\{v_k\}_{k=0}^\infty$\ , with the following action of the generators
\beq
\SpinAux^3\,v_k=(j-k)\,v_k,\qquad
\SpinAux^-\,v_k=\,v_{k+1},\qquad
\SpinAux^+\,v_k=k\,(2j-k+1)\,v_{k-1}\ .\label{pijplus}
\eeq
The convergence of the trace in \eqref{Tjplus} requires one to assume
that $\mbox{Im}\,\phi<0$.  For a generic value of $j\in{\mathbb C}$
the representation \eqref{pijplus} is irreducible. However when $j$
takes non-negative (half) integer values $2j\in{\mathbb Z}_{\ge0}$ this
representation becomes reducible. The 
matrices $\pi_j^+(J^+)$, $\pi_j^+(J^-)$  and 
$\pi_j^+(J^3)$\  then acquire a block-triangular form with two diagonal blocks.
One of these is finite-dimensional, being equivalent to the $(2j+1)$-dimensional
representation $\pi_j$, and the other one is infinite-dimensional, and coincides with the
highest weight representation $\pi_{-j-1}^+$. Hence on the level of
traces one easily obtains\footnote{%
By the methods of appendix \ref{app:finiteTj} the relation \eqref{tjtplus} may also be analytically continued to complex $j$, such that
$\Top_j(\specbaz)$ stays finite in the $\twist \rightarrow 0$ limit.
}
\begin{equation}
\label{tjtplus}
\Top_j(\specbaz) \equiv
  \Top^+_j(\specbaz) 
 - \Top^+_{-j-1}(\specbaz)\, ,\qquad  2j\in{\mathbb Z}_{\ge0}\ .
\end{equation}
%


\subsection{Functional relations}

Our immediate aim is to study various algebraic properties of the transfer
matrices \eqref{Tj} and \eqref{Tjplus}, and, in particular, to derive
all functional equations they satisfy.  In the context of integrable
models, related to the quantized Kac-Moody algebra
$U_q(\widehat{sl}(2))$, this problem has been previously solved in
\cite{Bazhanov:1996dr,Bazhanov:1998dq}. Here we will apply a similar approach. The detailed
considerations of \cite{Bazhanov:1996dr,Bazhanov:1998dq} are devoted to conformal field
theory and cannot be straightforwardly used for the lattice $XXX$-model.
Moreover, the latter model is related to a rather subtle limit
$q^2\to1$ in the relevant $q$-deformed constructions of \cite{Bazhanov:1996dr,Bazhanov:1998dq}, and requires additional considerations. Nevertheless one should expect that the
general structure of the functional relations, independent of
the value of the deformation parameter $q$, must remain intact in the 
$q^2\to1$ limit by continuity arguments. In particular, one should expect 
from Eq.(2.45) of 
\cite{Bazhanov:1996dr} that the 
transfer matrix $\Top^+_{j}(\specbaz)$, defined in \eqref{Tjplus}
above,
factorizes into a product 
\beq
\label{fact}
f(\twist)\, \Top^+_{j}(\specbaz)= \Qop_+(\specbaz+j+\half) \,
\Qop_-(\specbaz-j-\half)\, ,
\qquad {\rm where} \qquad
f(\phi)=2\,i\,\sin  \frac{\twist}{2}\, , 
 \end{equation}
of two Baxter ${\bf Q}$-operators $\Qop_\pm(z)$, which should satisfy
\beq
[{\bf T}_j(z),\Qop_\pm(z')]=0,
\qquad 
[\Qop_+ (z)\,,\Qop_-(z')]=0\, ,
\qquad 
[\Qop_\pm (z)\,,\Qop_\pm(z')]=0\, .\label{Qcomm}
\eeq
They clearly extend the commuting family \eqref{Tcomm}, and should satisfy 
the $\Top \Qop$-equation \eqref{OpBaxtereqQalt}. Below we will prove 
that this is indeed true in our case and give the corresponding
definitions of the operators $\Qop_\pm(z)$. 

Before going into this proof  let us demonstrate that
the relation \eqref{fact} alone leads to a simple derivation of {\em all}
functional relations, involving various ``fusion'' transfer matrices
${\bf T}_j(z)$ and $\Qop$-operators
\cite{Bazhanov:1996dr,Bazhanov:1998dq}. For this reason  
Eq.~\eqref{fact} can be regarded as  a {\em universal fusion
  relation} --- once it is derived, no further algebraic work is
required. 

Substituting \eqref{fact} into \eqref{tjtplus} one obtains\footnote{This equation appeared for the XXX model in \cite{Korff:2006}. However, there the Q-operators were not completely explicit {\em cit.} ``the operators $Q^\pm_\omega(0)$ are not easily determined and we are missing at the moment concrete expressions for them."
}
 \begin{equation}
 \label{fundrel}
 f(\twist)\, \Top_{j}(\specbaz)=
 \Qop_+(\specbaz+j+\half)\,\Qop_-(\specbaz-j-\half)  - 
 \Qop_-(\specbaz+j+\half)\, \Qop_+(\specbaz-j-\half)\, ,\quad 2j\in{\mathbb Z}_{\ge0}\, .
 \end{equation}
For $j=0$ (trivial representation) one clearly has
\begin{equation}\label{trivialtransfer}
\Top_0(\specbaz)=z^L\, \mathbb{I}\, ,
\end{equation}
where $\mathbb{I}$ is the identity operator on the quantum space, which hereafter will be omitted.
In this case \eqref{fundrel} reduces to 
 \begin{equation}
 \label{qwron}
  \Qop_+(\specbaz+\half)\,\Qop_-(\specbaz-\half)  - 
 \Qop_-(\specbaz+\half)\, \Qop_+(\specbaz-\half)=
f(\twist)\, z^L\ .
 \end{equation}
Using the last relation together with the expression \eqref{fundrel}
for ${\bf T}_{\frac{1}{2}}(z)\equiv{\bf T}(z)$ one immediately derives
the $\Top \Qop$-equation \eqref{OpBaxtereqQalt}.
Note that the relation
\eqref{qwron} can be regarded as the 
(quantum) Wronskian relation for the second
order difference equation \eqref{OpBaxtereqQalt}, ensuring linear
independence of the two solutions $\Qop_+$ and
$\Qop_-$. 
Furthermore, \eqref{fundrel} clearly 
indicates that the operators $\Qop_\pm$ should be considered to be
more fundamental than the transfer matrices, since  the latter are a
quadratic superposition of the former.  

In fact,  we can consider \eqref{fundrel} as the most fundamental
fusion relation of the model, from which all other relations follow. 
The mechanism by which this happens is quite simple.  Let us define,
for any commuting quantity $\Qop^{\pm}_A$, 
 \begin{equation}
\label{fundrelcompact}
 \mathcal{T}_{AB} \equiv \Qop^+_{[A}\, \Qop^-_{B]}\, ,
 \end{equation}
where the square brackets indicate antisymmetrization and the indices
$A,B$ denote very generally a collective set of discrete indices
and/or continuous variables. The latter are related to linear
combinations of the the spectral parameter $\specbaz$ and some
representation labels such as the spin $j$. 
We immediately infer the identities
 \begin{equation}
\label{funcrelcompact}
\mathcal{T}_{[AB}^{}\, \Qop_{C]}^{\pm}=0 \, , \qquad \qquad
\mathcal{T}_{[AB}\, \mathcal T_{C] D}=0\, .
 \end{equation}
These two types of equations are a compact way to respectively write 
Baxter's equation \eqref{OpBaxtereqQalt} as well as all fusion
relations \eqref{fusrel} such as the ones of \cite{Kirillov:1987zz}:
\beq\label{fusrel}
{\bf T}_j(z+\half)\,{\bf T}_j(z-\half)=\left(z+j+\half\right)^L\left( z-j-\half\right)^L+
{\bf T}_{j+\frac{1}{2}}(z)\,{\bf T}_{j-\frac{1}{2}}(z), \quad
j=0,\half,1,\ldots 
\eeq
A generalization of these relations appears in \Appref{app:functionalrel}. The same kind of generalized fusion relations can be found in \cite{Derkachov:2005xn}.


\subsection{Factorization of the \texorpdfstring{$\mathcal{L}$}{}-operator}

As noted before, we want to construct the ${\bf Q}$-operators as 
transfer matrices \eqref{tmat} built from suitable $L$-operators,
solving the Yang-Baxter equation \eqref{YB0}. We shall soon see that the
required $L$-operators indeed exist. They are easily obtained via some
special reductions of \eqref{sl2invlax}. In preparation of this
calculation we need to recall some well-known realizations of the
$\alg{sl}(2)$ commutation
relations \eqref{sl2commrel} in terms of the harmonic oscillator algebra
\beq
{\mathsf H}:\qquad [\osch,\osca^\pm]=\pm\osca^\pm,\qquad
[\osca^-,\osca^+]=1,\qquad
\osch=\osca^+\osca^-+{\textstyle\frac{1}{2}}\, .\label{harm} 
\eeq
The Fock representation
${\cal F}$ of
this algebra is spanned on the vectors  
$\{v_k\}_{k=0}^\infty $\ ,
\beq
{\cal F}: \qquad \osca^+\,v_k=v_{k+1},\qquad \osca^-\,v_k=k\,v_{k-1},
\qquad \osch\,v_k=(k+\half)\,v_k \ .\label{fock} 
\eeq
The value of the quadratic Casimir operator of the algebra $\alg{sl}(2)$ 
\begin{equation}
C_2 \equiv \vec \SpinAux^2=(\SpinAux^3)^2+ \half \left( 
\SpinAux^+\, \SpinAux^- + \SpinAux^-\, \SpinAux^+\right)
\end{equation}
for the highest weight representations \eqref{h-weight} is given by 
\beq
\pi_j(C_2)=\pi_j^+(C_2)=j(j+1)\ .
\eeq
Below we will use the fact that for the infinite-dimensional
representation $\pi_j^+$ 
the $\alg{sl}(2)$-generators
\beq
\SpinAux^a_j=\pi_j^+(\SpinAux^a),\qquad a=\{3,+,-\},
\eeq
can be realized through the oscillator algebra \eqref{harm}
\begin{equation}
\label{HPfirst}
\SpinAux_j^-= \osca^+  \, ,
 \qquad \SpinAux_j^+=\left( 2\,j-  \osca^+\, \osca^-  \right) \osca^- \, ,
 \quad \SpinAux_j^3= j- \osca^+\, \osca^- \, ,
\end{equation}
or alternatively as
\begin{equation}
\label{HPsecond}
\SpinAux_j^-= \osca^+ \left( 2\,j-  \osca^+\, \osca^-  \right)  \, ,
 \qquad \SpinAux_j^+= \osca^- \, ,
 \quad \SpinAux_j^3= j- \osca^+\, \osca^- \, ,
\end{equation}
where the operators $\osca^+,\osca^-$ are taken in the Fock
representation \eqref{fock}. These realizations are commonly known 
as Holstein-Primakoff representations;
they can be readily verified. Indeed, having in mind the Fock
representation \eqref{fock}, one can easily see 
that the formulae \eqref{HPfirst} are 
essentially verbatim transcriptions of the matrix elements
\eqref{pijplus} in the oscillator notations. The second realization
\eqref{HPsecond} is obtained from \eqref{HPfirst} by a simple similarity
transformation\footnote{%
One can also  take 
\begin{equation}
\SpinAux_j^-= \osca^+ \left( 2\,j-  \osca^+\, \osca^-  \right)^{\gamma} \, ,
 \qquad \SpinAux_j^+=\left( 2\,j-  \osca^+\, \osca^-  \right)^{1-\gamma} \osca^- \, , 
 \quad \SpinAux_j^3= j- \osca^+\, \osca^- \, .
\nonumber\end{equation}
with arbitrary $0\le\gamma\le1$. Here we only use the values $\gamma =
0,1$.}, and, therefore gives the same trace in
\eqref{Tjplus}\footnote{%
This is also true for $2j\in {\mathbb Z}_{\ge0}$,  when \eqref{HPfirst} and
\eqref{HPsecond} become reducible and on the level of traces both lead to
the same formula \eqref{tjtplus}}.

Due to the invariance of the $R$-matrix \eqref{gl2inv}, the solutions of
\eqref{YB0} are defined up to linear transformations  
\beq
\Lbb_V(z)\to F\,\, \Lbb_V(z)\, G,\qquad F,G\in {\rm GL}(2)\ ,\label{fgconj}
\eeq
where $F$ and $G$ are arbitrary non-degenerate $2\times2$ matrices.
In other words, the transformation \eqref{fgconj} does not affect the
validity of the Yang-Baxter equation \eqref{YB0}.  

Using the realization \eqref{HPfirst}, let us write explicitly the
$L$-operator \eqref{sl2invlax} in the representation $\pi_j^+$,
\beq
L_j(z)=\pi_j^+(L(z))=\left(\begin{array}{cc}
z+j-\osca^+\osca^-& \osca^+\\[.2cm]
(2j-\osca^+\osca^-)\,\osca^-&\qquad
z-j+\osca^+\osca^-\end{array}\right)\ .
\label{Ljmatrix}
\eeq 
Its matrix elements depend on two parameters $z$ and $j$. It is
convenient to define new variables 
\beq\label{lightconespec}
z_\pm=z\pm(j+\half)\ .
\eeq
Consider the limit
\beq
z\to\infty,\qquad j\to\infty, \qquad z_+=\mbox{fixed}\, ,
\eeq
and define  
\beq
L_+(z_+)=\lim_{j\to\infty}\left(\begin{array}{cc}1&\quad 0\\
0&\quad -\frac{1}{2j}\end{array}\right) L_j(z_+-j-\half) =
\left( \begin{array}{cc}
\specbaz_+-\osch \, \, & ~~\osca^+\, \, \\
-\osca^- \, \,&~~1\, \, \end{array} \right)\ ,\label{Lplus}
\eeq
where $\osch$ is defined in \eqref{harm}. 
Similarly, for fixed $z_-$ define 
\beq
L_-(z_-)=\lim_{j\to\infty} 
L_j(z_-+j+\half)\,\left(\begin{array}{cc}\frac{1}{2j}&\quad 0\\
0&\quad 1\end{array}\right) =
\left( \begin{array}{cc}
1 \ \  & ~\osca^+\, \, \\
\osca^- \ \  \,&z_-+\osch\end{array} \right)\ .\label{Lminus}
\eeq
The particular transformations of the form \eqref{fgconj}, 
used in the definitions
\eqref{Lplus} and \eqref{Lminus}, do not affect the validity of  
the Yang-Baxter equation \eqref{YB0}. Therefore, the new operators $L_\pm(z)$
will automatically satisfy \eqref{YB0} in
virtue of the commutation relations of the oscillator algebra.
Of course this may also be verified by direct, elementary calculations.
It should be mentioned that these two $L$-operators are not really 
new objects, and appeared earlier in different contexts. 
They are e.g.~known to yield the Lax operators of the so-called
``discrete self-trapping chain'' \cite{Kuznetsov:1999tk,kovalsky}.
 
A typical problem frequently arising in the theory of integrable
system is the ``fusion'' of different solutions of the Yang-Baxter
equation. This is a standard way to understand relationships between 
various solutions and to obtain new ones.  It turns out that in 
our case the consideration of the product of $L_+(z)$ and $L_-(z')$ 
does not lead to any new non-trivial solutions. It however allows to
discover remarkable factorization properties of the operator $L_j(z)$,
which will be explained below.

Consider the direct product ${\mathsf H}^{\otimes 2}$ 
of two oscillator algebras \eqref{harm}. 
Let $\{\osca^\pm_1, \osch_1\}$ and $\{\osca^\pm_2, \osch_2\}$ denote
the two associated mutually commuting sets of generators.
Below we will use the following similarity transformation acting in this
direct product  
\begin{equation}
\widetilde\osca_k^\pm = e^{\osca_\indone^+ \osca_\indtwo^-}\,
\osca_k^\pm\, e^{-\osca_\indone^+ \osca_\indtwo^-} \,  ,\qquad k=\indone,\indtwo.
\end{equation}
Obviously, the transformed operators $\widetilde\osca_{\indone,\indtwo}^\pm$ 
obey exactly the same commutation relations as
$\osca_{\indone,\indtwo}^\pm$. In particular, the operators
$\widetilde\osca^\pm_\indone$ commute with
$\widetilde\osca^\pm_\indtwo$. Explicitly one has,
\begin{equation}
\widetilde\osca_\indone^-=\osca_\indone^--\osca_\indtwo^- \,  ,
\qquad  \qquad
\widetilde\osca_\indone^+=\osca_\indone^+ \, ,
\end{equation}
\begin{equation}
\widetilde\osca_\indtwo^-=\osca_\indtwo^- \,  ,
\qquad  \qquad 
\widetilde\osca_\indtwo^+=\osca_\indone^++\osca_\indtwo^+ \, .
\end{equation}
Consider the product 
\beq
{\mathcal L}(z)=L^{(\indone)}_-(z_-)\, L^{(\indtwo)}_+(z_+)
=
\left( \begin{array}{cc}
1~~\, \,& \osca_\indone^+ \, \, \\
\osca_\indone^-~~\, \, &  \specbaz_- +\osch_\indone,
\,\end{array} \right)
\left( \begin{array}{cc}
\specbaz_+-\osch_\indtwo\, \, & ~~\osca_\indtwo^+\, \, \\
-\osca_\indtwo^- \, \,&~~1\, \, \end{array} \right) ,
\label{Lmp}
\end{equation}
where the superscripts $(\indone)$ or $(\indtwo)$ indicate that the corresponding 
$L$-operators belong respectively to the first or the second
oscillator algebra. 
By elementary calculations one can bring this product to the form
\begin{equation}
\label{firstway-final}
\left( \begin{array}{cc}
1~~\, \,& \osca_\indone^+ \, \, \\
\osca_\indone^-~~\, \, &  \specbaz_- +\osch_\indone,
\,\end{array} \right)
\left( \begin{array}{cc}
\specbaz_+-\osch_\indtwo\, \, & ~~\osca_\indtwo^+\, \, \\
-\osca_\indtwo^- \, \,&~~1\, \, \end{array} \right)
=
e^{\osca_\indone^+ \osca_\indtwo^-}\left( \begin{array}{cc}
1 \, \, & 0 \, \, \\
\osca_\indone^- \, \, & 1 \, \,  \end{array} \right)
\left( \begin{array}{cc}
\specbaz+ \SpinAux_j^3  \, \, &  \SpinAux_j^- \, \,  \\
\SpinAux_j^+ \, \, & \specbaz-\SpinAux_j^3  \, \, \end{array} \right)
e^{-\osca_\indone^+ \osca_\indtwo^-}\, .
\end{equation}
%
%
%
where $\osch_{1,2}=\osca^+_{1,2}\,\osca^-_{1,2}+\half$ and the $\alg{sl}(2)$ generators $\SpinAux^3_j$, $\SpinAux^+_j$ and $\SpinAux^-_j$ 
are realized as in (\ref{HPfirst}) but employing the operators 
$\osca^\pm_2$,
\begin{equation}
\label{HPforfact1-alt}
\SpinAux^-_j= \osca_\indtwo^+ \, , \qquad
 \SpinAux_j^+=\left(2\, j - \osca_\indtwo^+\, \osca_\indtwo^-   \right) \osca_\indtwo^-\, ,
 \quad \SpinAux^3_j= j- \osca_\indtwo^+\, \osca_\indtwo^-  \, .
\end{equation}
Introducing operator valued matrices
\begin{equation}\label{Bops}
 B_+= 
\left( \begin{array}{cc}
1\, & \osca^+ \, \\
0\,&1\, \end{array} \right)   \, , 
\qquad 
 B_-= 
 \left( \begin{array}{cc}
1 \,& 0 \, \\
\osca^-\, &1\, \end{array} \right)   \, , 
\end{equation}
one can re-write \eqref{firstway-final} in a compact form
\begin{equation}\label{compact1}
{\mathcal L}(z)=
L^{(\indone)}_-(z-j-\half)\, L^{(\indtwo)}_+(z+j+\half)=
e^{\osca^+_\indone \osca^-_\indtwo}\ B_-^{(\indone)}\  
L_j^{(\indtwo)}(z)\ e^{-\osca^+_\indone
  \osca^-_\indtwo} 
\, ,
\end{equation}
where the superscripts $(\indone)$ and $(\indtwo)$ have the same meaning as in \eqref{Lmp}.
Similarly one obtains 
\begin{equation}\label{compact2}
\overline{{\mathcal L}}(z)=
L^{(\indone)}_+(z+j+\half)\, L^{(\indtwo)}_-(z-j-\half)=
e^{\osca^+_\indone \osca^-_\indtwo}\  
L_j^{(\indone)}(z)\ B_+^{(\indtwo)}\ e^{-\osca^+_\indone
  \osca^-_\indtwo} 
\, ,
\end{equation}
Writing the last identity in full, one gets
\begin{equation}
\label{secondway-final}
\left( \begin{array}{cc}
\specLbar-\osch_\indone \, \, &  ~~\osca_\indone^+ \, \, \\
-\osca_\indone^- \, \, &~~1 \, \, \end{array} \right)
\left( \begin{array}{cc}
1~~  \, \, & \osca_\indtwo^+  \, \, \\
\osca_\indtwo^-~~  \, \, & \specL+\osch_\indtwo \,
\, \end{array} \right)= 
 e^{\osca_\indone^+ \osca_\indtwo^-} \left( \begin{array}{cc}
\specbaz+ \SpinAux_j^3 \, \, &  \SpinAux_j^- \, \, \\
\SpinAux_j^+ \, \, & \specbaz-\SpinAux_j  \, \, \end{array} \right)
\left( \begin{array}{cc}
1 \, \, & \osca_\indtwo^+ \, \,  \\
0 \, \, & 1 \, \, \end{array} \right) e^{-\osca_\indone^+
  \osca_\indtwo^-}\,  ,
\end{equation}
where the $\alg{sl}(2)$ algebra is now realized as in
(\ref{HPsecond}) with  $\osca_\indone^\pm$ 
\begin{equation}
\label{HPforfact2-alt}
 \qquad \SpinAux_j^-=\osca_\indone^+
 \left(2\,j-\osca_\indone^+\,\osca_\indone^-  \right)\, ,\quad
 \SpinAux_j^+= \osca_\indone^-\, ,
\quad \SpinAux_j^3= j - \osca_\indone^+\, \osca_\indone^-   \, .
\end{equation}
%

 \subsection{Construction of the \texorpdfstring{${\bf Q}$}{}-operators}
 \label{sec:traces}

We are now ready to explicily define the operators ${\bf Q}_\pm(z)$ as
transfer matrices. To do this we will  use our general definition
\eqref{tmat} of a transfer matrix for the $L$-operators $L_\pm(z)$
defined in \eqref{Lplus} and \eqref{Lminus}. Solving \eqref{twistop}
one gets the same boundary operator in both cases,
\beq
\Dbb_\pm=e^{-i \phi\, \osch},
\label{Dpm}
\eeq
where $\osch$ is defined in \eqref{harm}. In this way we define
operators  
\beq
{\mathbf Q}_\pm(z)=Z^{-1}\,e^{\pm \frac{i}{2}{\phi}\,z}\,\Tr_{\mathcal F}\,\big(\Mcal_\pm(z)\big),      
\label{Qpm} \eeq
where 
\beq
\Mcal_\pm(z)=e^{-i\phi \/\osch}\underbrace{
{L}_\pm(z)\otimes L(z)_\pm\otimes \cdots \otimes  
L_\pm(z)}_{L-\mbox{\scriptsize{times}}}\ .
\eeq
Note that we have changed the normalization of \eqref{Qpm} 
in comparison with \eqref{tmat}
by introducing $z$-dependent exponents and a constant factor 
\beq
Z(\phi)=\Tr_{\mathcal F}\big( e^{-i\phi\, \/\osch}\big) =\frac{1}{2i \sin\frac{\twist}{2}}
\ .
\eeq 
where the subscript $\mathcal F$  indicates that the traces are taken 
over Fock space \eqref{fock}. 
By construction the operators \eqref{Qpm} will
automatically commute with ${\bf T}(z)$ as a consequence of the
Yang-Baxter equation \eqref{YB0} and properties
\eqref{gl2inv} and \eqref{twistop},
\beq
[{\bf T}(z),{\bf Q}_\pm(z')]=0\, .
\eeq
 
The operators ${\bf Q}_\pm(z)$ can be obtained from each other by
negation of the twist $\phi$. It is not difficult to show that
\begin{equation}
\Qop_+(\specbaz,\twist)={\mathcal R}\
\Qop_-(\specbaz,-\twist)\ {\mathcal R}\, , \qquad {\mathcal
  R}=\sigma_x\otimes
\sigma_x\otimes\cdots\otimes\sigma_x\, ,
\end{equation}
where ${\mathcal R}$ is the spin reversal operator in the quantum space.

Let us now prove the factorization equation \eqref{fact}. Consider the
quantity defined by the first equality in \eqref{compact1},
\beq
{\mathcal L}(z)=
L^{(1)}_-(z-j-\half)\, L^{(2)}_+(z+j+\half)\ .\label{Lcal2}
\eeq
where superscripts $(1)$ and $(2)$ have the same meaning as in \eqref{Lmp}.
This is a product of two $L$-operators with different auxiliary
spaces (each being a copy of the Fock space), which, of course, satisfies the
Yang-Baxter equation itself. It can be regarded as a single $L$-operator,
whose auxiliary space is the tensor product of two Fock spaces. Therefore,
one can apply our general formula \eqref{tmat} to define the transfer
matrix
\beq
\Tbb_{\mathcal L}(z)=
\Tr_{\mathcal{F}_1\mathcal{F}_2}\,\Big(e^{- i \twist\, (\osch_1+\osch_2)} 
{}\underbrace{
{\mathcal L}(z)\otimes {\mathcal L}(z)\otimes \cdots \otimes  
{\mathcal L}(z)}_{L-\mbox{\scriptsize{times}}}\Big)\, ,
\label{TLcal}
\eeq 
where we have substituted $\Dbb_\pm$ defined in \eqref{Dpm}. 
It is not difficult to see
that using \eqref{Lcal2} one can rearrange factors under the trace such
that \eqref{TLcal} reduces to a product of two operators \eqref{Qpm},
\beq
\Tbb_{\mathcal L}(z)=Z(\phi)^2 \,e^{-i \phi\,(j+\half)}\, {\bf Q}_-(z-j-\half)\,
{\bf Q}_+(z+j+\half).
\eeq
Similarly, using the second expression for ${\mathcal L}(z)$ from
\eqref{compact1} one obtains 
\beq
\Tbb_{\mathcal L}(z)= e^{-i \phi\,(j+\half)}\,\Tbb_{B_-}\,{\bf T}^+_j(z) ,
\eeq
where 
\beq
\Tbb_{B_-}=\Tr_{\mathcal{F}}\,\Big(e^{- i \twist\, \osch} 
{}\underbrace{
{B_-}\otimes {B_-}\otimes \cdots \otimes  
{B_-}}_{L-\mbox{\scriptsize{times}}}\Big).
\label{TB}
\eeq
Note that the presence of the similarity transformation in \eqref{compact1}
does not affect the calculation of the trace since the operator
$e^{\osca_\indone^+ \osca_\indtwo^-}$ commutes with $\osch_1+\osch_2$ in
the boundary twist. The matrix $B_-$, defined in \eqref{Bops}, is
triangular. The calculation of the trace in \eqref{TB} is trivial, since $B_-$ depends only on $\osca^-$, it gives
\beq
\Tbb_{B_-}=Z(\phi)\, {\mathbb I}\ ,
\eeq
where ${\mathbb I}$ is the unit operator in the quantum space. 
Combining everything, one arrives at the factorization formula
\eqref{fact}, though with swapped order of ${\bf Q}_+$ and ${\bf Q}_-$ 
in the product.
Repeating the same reasonings, but this time starting from the equation
\eqref{compact2}, one obtains \eqref{fact} exactly as written,
which proves the commutativity of ${\bf Q}_+$ and ${\bf Q}_-$, stated
in \eqref{Qcomm}.

Let us stress that one should explicitly show that also the last equation in \eqref{Qcomm} 
is satisfied by the  operators constructed above. This  can be proven in the usual way
 starting from the Yang-Baxter equation in $\mathbb{C}^2 \otimes osc \otimes osc$.
The existence of an intertwiner in this case has been shown in \cite{kovalsky} in the course of a study of the DST chain, which is a certain bosonic hopping model. The further investigation of this and related issues will be reported elsewhere \cite{blms}.

For the benefit of the reader, we will present for small chain lengths $L=1,2$ the explicit forms of the finite twist $\twist$ operators $\Top$, $\Qop_\pm$ as well as their eigenvalues in appendix \ref{app:smalllengths}.

\subsection{Bethe equations without a Bethe Ansatz}
\label{sec:bethe}
It is interesting that our factorization formulas
\eqref{firstway-final} and \eqref{secondway-final} allow to solve the
twisted Heisenberg model without the somewhat tedious Bethe Ansatz
technique, be it a coordinate, algebraic or functional Bethe
``Ansatz''! In fact, no Ansatz (German for ``try and see whether it
works'') for some wavefunction is ever made. One instead derives the
fundamental operator relation \eqref{fundrel} as done in the last
section.  The full hierarchy of operatorial Baxter and fusion
equations immediately follows, as we explained in the course of the discussion of
\eqref{funcrelcompact}. In particular, one immediately derives the
operatorial Baxter equation \eqref{OpBaxtereqQalt} from \eqref{fundrel}
with auxiliary spin $j=\half$. 

As shown above all ${\bf T}$- and ${\bf Q}$-operators belong to the
same commuting family and therefore can be simultaneously diagonalized by 
a $z$-independent similarity transformation. Thus the eigenvalues will
have the same analytic properties in the variable $z$  as the matrix elements of the corresponding operators. 
By construction all ${\bf T}$-operators are polynomials in $z$, 
while the ${\bf Q}$-operators are polynomials multiplied by the simple 
$z$-dependent exponents (see \eqref{Qpm}) so that their eigenvalues
are exactly of the form \eqref{baxterfunction}, \eqref{baxterpoly}. 
Furthermore,
all the functional equations can be considered in the basis where all
operators are diagonal and can be replaced by their eigenvalues
corresponding to the same eigenstate. In other words the functional
equations can be treated as scalar equations for the eigenvalues.
For instance, substituting \eqref{baxterfunction}, \eqref{baxterpoly} into 
\eqref{qwron} one obtains 
\begin{equation}
 \label{qwron2}
 e^{\frac{i}{2}\phi} A_+(\specbaz+\half)\,A_-(\specbaz-\half)  - 
  e^{-\frac{i}{2}\phi} A_-(\specbaz+\half)\, A_+(\specbaz-\half)=
f(\twist)\, z^L\ .
 \end{equation}
Let $z_k$ be a zero of $A_+(z)$, i.e, $A_+(z_k)=0$. Using this fact in the last equation 
one obtains,
\beq
e^{-\frac{i}{2}\phi} A_-(\specbaz_k)\,
A_+(\specbaz_k-1)=(z_k-\half)^L,\quad
e^{\frac{i}{2}\phi} A_+(\specbaz_k+1)\,A_-(\specbaz_k)= 
(z_k+\half)^L,
\eeq
Dividing these equations by each other one arrives at
one set of the Bethe equations \eqref{betheeqsfromA}. The other set is
derived similarly.

We would like to mention one simple but important corollary of the quantum Wronskian relation 
\eqref{qwron2}. It concerns {\em exact strings}, i.e.,
groups of zeroes $\{z_1,z_2,\ldots,z_\ell\}$ 
equidistantly spaced  with the interval $1$,
\beq
z_{k+1}=z_k+1,\qquad k=1,\ldots,\ell-1,
\eeq
where
$\ell$ is called the length of the string.
It is obvious from \eqref{qwron2} that 
neither of the polynomials $A_\pm(z)$ could have {\em exact strings}
of the length greater then two. Indeed, the only such zeroes can be at 
$z=\pm\half$, thus $A_\pm(z)$ might only have (possibly multiple) 2-strings 
at $z=\pm\half$.  The same analysis extends to the zero-field case as
well. Note, in particular, that the long strings on the imaginary axis
of the variable $u=iz$, shown in
fig.~\ref{L60M22} in section \ref{sec:numerics}, are not exact.

\section{Removing the Twist: The \texorpdfstring{$\phi \rightarrow 0$}{} Limit}
\label{sec:notwist}

In the previous chapter \ref{sec:construction} we have used a regulator $\twist$ to obtain finite 
quantities when taking traces over infinite dimensional oscillator spaces. It is interesting that this natural quantity $\twist$ corresponds precisely to the spin chain with magnetic flux discussed in the review chapter \ref{sec:review}. In the present chapter we will deal with the operation of taking this regulating flux away. This involves a significant symmetry enhancement to a global SU(2) symmetry of the spectrum of the Heisenberg chain, ``weakly'' broken by the twist. Moreover we would like to make contact with the results of \cite{Pronko:1998xa}, see also the attempt in \cite{Pronko:1999is}, as well as the findings in \cite{Korff:2006}. 
It is interesting to explicitly work out the operators  $\Qop_\pm$ for small spin chain lengths $L$. This is done in appendix \ref{app:smalllengths}. In line with their construction, the operators $\Qop_\pm$ will diverge as the regulating $\twist$ is removed, and we will need to ``renormalize'' them. The reason is simply that the sum over oscillator states ceases to converge, as maybe explicitly verified on the examples in appendix \ref{app:smalllengths}. Let us begin by studying the structure of divergencies of $\Top^+_j$ and $\Qop_\pm$.

\subsection{Divergencies in the zero twist limit} 
\label{sec:zerotxistdiv}

It is interesting to understand the singularity structure\footnote{
We do not have proofs for the singular behavior of the operators $\Top^+$ and $\Qop^{\pm}$  discussed in the following. Our statements are based on observations on spin chains of small length $L=2,\ldots,5$.
}
 of $\Top,  \Top^+$ and $\Qop^{\pm}$ in the limit where the twist $\twist$ is sent to zero, and $\alg{su}(2)$ invariance is recovered.
In preparation, we will prove in appendix \ref{app:finiteTj} that $\Top_j(\specbaz; \twist)$
is finite in the zero-twist limit for arbitrary complex values of  $\specbaz$ and $j$, which is a priori not obvious. It is trivially finite for  $2j \in \mathbb{Z}^+$, since in this case there is near complete cancellation between two infinite sums, the remainder being a sum over $2\,j+1$ terms.  As for $\Top^+_j(\specbaz; \twist)$, one can immediately say that its eigenvalues diverge\footnote{
Note that
\begin{equation}
\sum_{n=0}^\infty e^{-xn}\, n^l \simeq (l-1)!\, x^{-(l+1)}
\end{equation}
in the $x \rightarrow 0$ limit.}
at most as $\twist^{-L}$ in the zero-twist limit. However, we found evidence that the actual divergence is much milder, namely\footnote{
It is not clear to us whether $\Op(\specbaz,j)$ can be meaningfully interpreted as a ``renormalized'' transfer matrix.
E.g.~we do not know whether this operator is $\alg{su}(2)$ invariant. An alternative, more symmetric definition would be
 $ \Top^+_j(\specbaz; \twist)$  $ \rightarrow$ $ \twist^{-\mathcal J}$ $\Op (\specbaz,j) $ $ \twist^{-\mathcal J}$.
 }
 \begin{equation}
 \Top^+_j(\specbaz; \twist)  \rightarrow \twist^{-2 \mathcal J} \Op(\specbaz,j)\, ,
 \end{equation}
with some matrix $\Op(\specbaz,j)$. Here $\mathcal J$ is defined as
\begin{equation}
\mathcal J \left( \mathcal J +1 \right)= \vec \SpinQ_{tot} ^2\, .
\end{equation}
This being a quadratic equation, there are two solutions for $\mathcal J$, and one has to choose the one such that the spectrum of  $2\mathcal J$ belongs to $ \mathbb{Z}_{\geq 0}$.

In any case, we already showed that the operator  $\Top_j^+(\specbaz)$ is a ``composite object'', and thus perhaps not too interesting. Much more relevant is the zero-twist limit of its fundamental constituents $\Qop_\pm (\specbaz_\pm)$.
As we discussed in the review chapter \ref{sec:review}, we do expect these operators to become singular, as Bethe roots at infinity physically indicate $\alg{su}(2)$ descendent states. Their naive divergent behavior is $\twist^{-(\frac{L}{2}\pm \SpinQ^3)}$, but their actual divergence is again empirically found to be much milder
 \begin{equation}
 \Qop_\pm(\specbaz_\pm; \twist)  \rightarrow \twist^{- (\mathcal J \pm \SpinQ^3)} \Op^{\pm}(\specbaz_\pm)\, .
 \end{equation}
In particular their action on lowest and highest weight states is finite in the zero-twist limit for $\Qop^+$ and $\Qop^-$, respectively.
One can check on small-length examples that $\Op^{\pm}(\specbaz)$  defined in this way is not $\alg{su}(2)$ invariant.
An $\alg{su}(2)$ invariant operator $\Qop(\specbaz)$ may be defined as
\begin{equation}
\label{QfromQpmJ}
\Qop(\specbaz)= \lim_{\twist \rightarrow 0} \frac{(\mathcal J \mp \SpinQ^3)!}{(2\mathcal J )!} 
(\mp \twist)^{\frac{1}{2} (\mathcal J \pm \SpinQ^3)}
\Qop_{\pm}(\specbaz;\twist)
(\mp \twist)^{\frac{1}{2} (\mathcal J \pm \SpinQ^3)}\, ,
\end{equation}
where it may be observed that the eigenvalues of $\Qop$ are polynomials in $\specbaz$ normalized as $\specbaz^{2 \mathcal J}+ \dots$. We see that both $\Qop_{\pm}$ lead to the {\it same} Baxter operator $\Qop$, see also \cite{Korff:2006}. This is technically easy to understand since $\Qop_{\pm}$ are related to each other by a spin-flip combined with reversion of the twist $\twist \rightarrow -\twist$. An $\alg{su}(2)$
invariant operator like $\Qop$ at $\twist=0$ is obviously invariant under this transformation. However, this means that in the limit we are ``loosing" one of the two linearly independent operators of the twisted case.


\subsection{\texorpdfstring{$\Qop$}{}-operator in the zero twist limit}
\label{sec:Qopzerotwist}

In this section we will propose another way to define the $\Qop$-operator in the zero-twist limit. 
This second way is certainly less explicit than \eqref{QfromQpmJ}, but is more suitable for rigorously proving finiteness, as well as $\alg{su}(2)$-invariance of the resulting $\Qop$-operator. The renormalized $\Qop(z)$ can be very naturally defined by ``renormalizing'' $\Qop_{\pm}(\specbaz)$, namely\footnote{%
Notice that $\Qop_{z_0}(z)= \Qop(z_0)^{-1}\Qop(z)$. Something similar appears
in  Baxter's original work \cite{Baxter:1972hz}. In \cite{Baxter:1972hz}  Baxter built two solutions to his equation, termed
$\Qop_L, \Qop_R$. These two matrices do not commute with the $\Top$ matrix and do not commute among themselves 
(they satisfy $\Qop_L(u)\Qop_R(v)=\Qop_L(v)\Qop_R(u)$). The operator $\Qop(u)\equiv\Qop_R(u) \Qop_R^{-1}(u_0)=\Qop_L(u) \Qop_L^{-1}(u_0)$
 is introduced to cure these two problems. We find this similarity quite interesting.
 }
\begin{equation}
\label{QfromQpmInversion}
\Qop_{\specsubtraction}(\specbaz)\equiv
 \lim_{\twist \rightarrow 0}
 \Qop_{\pm}(\specbaz;\twist) \,
 \Qop_{\pm}^{-1}(\specsubtraction;\twist).
\end{equation}
The resulting $\Qop$ still carries a dependence on the ``subtraction point'' $\specsubtraction$.
In this sense \eqref{QfromQpmInversion} defines a one parameter family of $\Qop$-operators.
$\Qop_{\specsubtraction}(\specbaz)$ in \eqref{QfromQpmInversion} satisfies Baxter's equation.
This is easily seen since Baxter's equation is linear in $\Qop$, and, at fixed values of $\twist$,
$\Top_j(\specbaz; \twist)$ and $\Qop_\pm(\specbaz; \twist)$ commute among themselves.
It is important to stress that the $\Qop$-operators defined in \eqref{QfromQpmInversion}
are $\alg{su}(2)$ invariant. The proof of this fact is left to \Appref{app:sl2inv}.
$\alg{su}(2)$ invariance is the underlying reason why in \eqref{QfromQpmInversion} both $\Qop_+$
and $\Qop_-$ lead to the same $\Qop$. 
Let us also point out that the obtained $\Qop$-operator will turn into the identity operators for $\specbaz=\specsubtraction$, an observation which will be used later.

The fact that \eqref{QfromQpmInversion} is a finite quantity is easy to understand.
Indeed one can safely discuss the finiteness issue at the level of eigenvalues. The similarity transform that diagonalizes $\Qop_{\pm}$ is well defined for any $\twist$ (including $\twist=0$), being the same similarity transform that diagonalizes $\Top_j(\specbaz; \twist)$.
Any given (fixed) eigenvalue of $\Qop_\pm (\specbaz)$ will be of the form
\begin{equation}
q_{\pm}(\specbaz, \twist)= \twist^{-k} \left( \, \tilde{q}_{\pm}(\specbaz) +\mathcal{O} (\twist)\, \right) ,
\end{equation}
for some  integer $k$.
The ratio 
\begin{equation}
\label{ratioqeigs} 
\frac{q_\pm (\specbaz; \twist)}{q_\pm (\specsubtraction;\twist)}\, ,
\end{equation}
will be finite in  the limit. Of course \eqref{QfromQpmInversion}
requires $\det \Qop_{\pm}(\specsubtraction;\twist) \neq 0$. To be precise, $\specsubtraction$ should not be a Bethe root for all values of $\phi$ in a small open domain in the vicinity of $\phi=0$. The zero twist limit of \eqref{ratioqeigs} is a polynomial in $\specbaz$ 
normalized to be one for $z=\specsubtraction$.
The degree of this polynomial can be worked out using $\alg{su}(2)$ invariance
of $\Qop_{\specsubtraction}(\specbaz)$ and the fact that $\Qop_+$ and $\Qop_-$ are finite
(because the Bethe roots stay finite) in the zero twist limit for, respectively, lowest and highest weight states. 
Moreover notice that the exponential in  \eqref{baxterfunction} can be simply replaced by one 
in the limit \eqref{QfromQpmInversion}. 

We finish this section by noticing that any two $\Qop_{\specsubtraction}(\specbaz)$ defined in \eqref{QfromQpmInversion}
are linearly dependent since
\begin{equation}
\label{Qoplindep}
\Qop_{\specsubtraction}(\specbaz_1)\, \Qop_{\specsubtraction'}(\specbaz_2) -\Qop_{\specsubtraction}(\specbaz_2)\, \Qop_{\specsubtraction'}(\specbaz_1) =0 \, .
\end{equation}
In the next section we will discuss how to obtain another one-parameter family of solutions to Baxter's equation, which will turn out to be linearly independent to the ones we just obtained in \eqref{QfromQpmInversion}.


\subsection{\texorpdfstring{$\Pop$}{}-operator (as a \texorpdfstring{$\Top$}{}-operator) in the zero twist limit.}
\label{sec:Popzerotwist}

In the previous section we identified a one-parameter family $\Qop_{\specsubtraction}(\specbaz)$ of solutions to the Baxter equation
in the zero twist limit. As already discussed in section \ref{sec:review}, Baxter's equation possesses two linearly independent solutions.
Accordingly, let us now obtain a second, linearly independent family of solutions, which we will denote by
$\Pop_{\specsubtraction}(\specbaz)$.
While constructed in a different way, it is identical to the one obtained in \cite{Pronko:1999is}, and is based on the same observation\footnote{
To avoid confusion, we note 
that our $\Pop$  was called $\Qop$ in \cite{Pronko:1999is}. However, its eigenvalues coincide, up to overall state-dependent normalization, with the polynomials $P$ in \cite{Pronko:1998xa}.}.
We will relate this second one-parameter family of operators to $\Qop_\pm(\specbaz; \twist)$.
The observation made in \cite{Pronko:1999is} is that 
\begin{equation}
\label{eq:oneparameterPfromT}
 \Pop_{\specsubtraction}(\specbaz)  \equiv \Top_{\sfrac{\specsubtraction-\specbaz-1}{2}} (\sfrac{\specsubtraction+\specbaz}{2}) \, ,
\end{equation}
satisfies Baxter's equation. 
$\specsubtraction$ is then the label of the one-parameter family.
 Notice that in this equation there is no twist and $\Pop_{\specsubtraction}(\specbaz) $ is an $\alg{su}(2)$ invariant operator. 

For $ \Pop_{\specsubtraction}(\specbaz)$  to be defined for any complex $\specbaz$, the $\Top$-operator has to be constructed for any complex value of the spin.
Different approaches to this problem are reviewed in  \Appref{app:finiteTj}. They include the construction proposed in  \cite{Pronko:1999is},
 the \emph{trace functional} approach introduced in  \cite{Boos:2004zq} and the approach of the present paper (see also \cite{Korff:2006}). 

It is instructive to explicitly write \eqref{eq:oneparameterPfromT} using the fundamental relation \eqref{fundrel}
\begin{equation}
\label{PopzerotwistfromQpm}
  \Pop_{\specsubtraction}(\specbaz)    \equiv \lim_{\twist \rightarrow 0}
 f^{-1}(\twist) \left(\Qop_-(\specbaz;\twist)\, \Qop_+(\specsubtraction;\twist) - 
 \Qop_-(\specsubtraction;\twist)\, \Qop_+(\specbaz;\twist) \right),
\end{equation}
and to take a closer look at the mechanism by which the eigenvalues of $\Pop_{\specsubtraction}(\specbaz)$ become
polynomials of the degrees found in \cite{Pronko:1998xa}.
A proof  is encoded in the zero twist limit of the  generalized Wronskian equation analyzed in the next section,
here we will merely give some heuristic argument.
In contradistinction to the $\Qop_{\specsubtraction}(\specbaz)$-operator, the $ \Pop_{\specsubtraction}(\specbaz)$-operator
has been defined as the difference of two quantities. Finiteness of \eqref{PopzerotwistfromQpm} is then a consequence 
of the cancellation of many divergent contributions from the two terms. This means that in the $\twist$ expansion
of \eqref{PopzerotwistfromQpm} the exponential factors appearing in \eqref{Qpm} and \eqref{baxterfunction} play an important role.
Their major role is to increase the degree of polynomiality of the $\Pop$-operator as compared to the naively expected one.

As opposed to the one-parameter family of $\Qop$-operators, 
any two $\Pop$-operators at different values of the parameter $\specsubtraction$ are linearly independent.
This is the underlying reason behind the appearence of an extra root in the $\Pop$-operator's
eigenvalues: in the zero twist limit the $\Pop$-operator contains the $\Qop$-operator but not vice versa.
Let us next analyze this in more detail.


\subsection{Generalized Wronskian for \texorpdfstring{$P$}{} and \texorpdfstring{$\Qop$}{}}

Thanks to the insights gained in the previous two sections, we will now study the relation between the  $\Qop$-operators and  $\Pop$-operators given 
in \eqref{QfromQpmInversion} and \eqref{PopzerotwistfromQpm}.
It is interesting to notice that they can be consistently defined for any value of the twist $\twist$.
Inspired by \eqref{QfromQpmInversion}, \eqref{PopzerotwistfromQpm}, let us define
\begin{equation}
\label{Ptauz}
\Pop_{\specsubtraction}(\specbaz;\twist)\equiv f^{-1}(\twist) \left(\Qop_-(\specbaz;\twist)\, \Qop_+(\specsubtraction;\twist)-
\Qop_+(\specbaz;\twist)\,\Qop_-(\specsubtraction;\twist) \right)\, ,
\end{equation}
\begin{equation}
\label{Qtauz}
\Qop_{\specsubtraction}(\specbaz;\twist)\equiv \frac{1}{2} \left(\Qop_-(\specbaz;\twist)\, \Qop_-^{-1}(\specsubtraction;\twist)+
\Qop_+(\specbaz;\twist)\,\Qop_+^{-1}(\specsubtraction;\twist) \right))\, .
\end{equation}
The explicit form of $f(\twist)$ is not important here, the important thing is that it is the same function
that appears in the fundamental relation.
As already shown,  both $\Qop_{\specsubtraction}(\specbaz;\twist)$ and $\Pop_{\specsubtraction}(\specbaz;\twist)$ satisfy  Baxter's equation\footnote{
Baxter's equation is the same for all these operators and does not contain the twist explicitly.
Its information is hidden in the analytic structure of the solution we want to obtain.
}
 and are finite in the zero twist limit. Here we want to stress that they
satisfy generalized Wronskian relations.
It can be shown by direct substitution, using the fundamental relation \eqref{fundrel}, that 
\begin{equation}
\label{TopjPQfinal}
\Top_{j}(\specbaz;\twist)= \Pop_{\specsubtraction}(\specbaz+j+\half;\twist)\,\Qop_{\specsubtraction}(\specbaz-j-\half;\twist)-
\Pop_{\specsubtraction}(\specbaz-j-\half;\twist)\,\Qop_{\specsubtraction}(\specbaz+j+\half;\twist)\,. 
\end{equation}

A question that naturally arises is the meaning of the parameter $\specsubtraction$.
It is interesting to notice that $\Top_{j}(\specbaz;\twist)$ in 
\eqref{TopjPQfinal} does not depend on the choice of $\specsubtraction$.
On the other hand, in the very same way as \eqref{fundrelcompact}, \eqref{fundrel} have an intrinsic $\alg{sl}(2)$ 
invariance\footnote{Note that this $\alg{sl}(2)$ is {\it not} identical to the manifest $\alg{su}(2)$ of our 
Heisenberg magnet under study. } 
which rotates $\Qop_\pm$ as a doublet,
$\Top_{j}(\specbaz;\twist)$ in \eqref{TopjPQfinal}  is left unchanged 
under rotations\footnote{In \Appref{app:moresl2wronskian} we review how  $\Qop_{\specsubtraction}(\specbaz;\twist)$ and $\Pop_{\specsubtraction}(\specbaz;\twist)$ transform 
under $\alg{sl}(2)$ rotations of $\Qop_\pm$ as a doublet. The fact that a $\alg{u}(1) \otimes \mathbb{Z}_2$ inside $\alg{sl}(2)$ acts freely on 
 $\Qop_{\specsubtraction}(\specbaz;\twist)$ and $\Pop_{\specsubtraction}(\specbaz;\twist)$ results in equations \eqref{Ptauz} and  \eqref{Qtauz}
not being invertible for $\Qop_\pm$.}
 of  $\Qop_{\specsubtraction}(\specbaz;\twist)$ and $\Pop_{\specsubtraction}(\specbaz;\twist)$.
 It turns out that these two invariances of \eqref{TopjPQfinal}
are not unrelated. It is easy to see that
\begin{equation}
\label{sl2tautaufreedom}
\begin{pmatrix}
\Pop_{\specsubtraction'}(\specbaz;\twist) \\
\Qop_{\specsubtraction'}(\specbaz;\twist)
\end{pmatrix} =
\begin{pmatrix}
\hat \alpha \,& \hat \beta\, \\
\hat \gamma \,& \hat \delta\,
\end{pmatrix}
\begin{pmatrix}
\Pop_{\specsubtraction}(\specbaz;\twist) \\
\Qop_{\specsubtraction}(\specbaz;\twist)
\end{pmatrix}=
\begin{pmatrix}
\Qop_{\specsubtraction}(\specsubtraction';\twist) \,&  \Pop_{\specsubtraction'}(\specsubtraction;\twist)\, \\
\Nop_{\specsubtraction'}(\specsubtraction; \twist) \,& \Qop_{\specsubtraction'}(\specsubtraction;\twist) \,
\end{pmatrix}
\begin{pmatrix}
\Pop_{\specsubtraction}(\specbaz;\twist) \\
\Qop_{\specsubtraction}(\specbaz;\twist)
\end{pmatrix}\, ,
\end{equation}
where
\begin{equation}
\label{lowertriangsl2}
\Nop_{\specsubtraction'}(\specsubtraction; \twist)=\frac{f(\twist)}{4}\left(\frac{1}{\Qop_-(\specsubtraction';\twist)\,
\Qop_+(\specsubtraction;\twist)}-\frac{1}{\Qop_-(\specsubtraction;\twist)\,\Qop_+(\specsubtraction';\twist)} \right) \, .
\end{equation}
We have  $\hat \alpha\, \hat \delta - \hat \beta\, \hat \gamma=1$, which is consistent with the $\alg{sl}(2)$ rotation.
This identity is a special case of the following interesting relation
\begin{equation}
\Nop_{\specsubtraction}(\specsubtraction';\twist) \Pop_{\specbaz'}(\specbaz;\twist)= \Qop_{\specsubtraction}(\specbaz;\twist)\,\Qop_{\specsubtraction'}(\specbaz';\twist)-
\Qop_{\specsubtraction}(\specbaz';\twist)\,\Qop_{\specsubtraction'}(\specbaz;\twist)\,. 
\end{equation}
Note that $\Nop_{\specsubtraction}(\specsubtraction';\twist)$ vanishes in the zero twist limit, in line with equation \eqref{Qoplindep}.
Let us stress the implications of this fact from the point of view of equation \eqref{sl2tautaufreedom}.
In the zero twist limit varying $\specsubtraction$ corresponds to triangular rotations of $\Pop$ and $\Qop$.
In particular, it means that the $\specsubtraction$ freedom corresponds to  the normalization of $\Qop$, but does not affect its Bethe roots.
For $\Pop$ the freedom in the parameter $\specsubtraction$ corresponds to adding the $\Qop$-operator to the $\Pop$-operator.
From the point of view of the Bethe roots distribution it corresponds to the position of one of the roots of $\Pop$.
It is indeed obvious from equation \eqref{Ptauz} that
$\Pop_{\specsubtraction}(\specsubtraction;\twist)=0$. The positions of the other Bethe roots then depend parametrically on the position of the root $\specsubtraction$. Note that the eigenvalues of \eqref{Ptauz} and \eqref{Qtauz} for the highest weight state in the zero twist limit will be

\begin{equation}
Q_{\specsubtraction}(\specbaz)=\prod_{k=1}^{M} \frac{\specbaz-\specbaz_k}{\specsubtraction-\specbaz_k}\, , \qquad
P_{\specsubtraction}(\specbaz)=(2 M + 1) \prod_{k=1}^{M}\left(\specsubtraction-\specbaz_k\right) \prod_{k=0}^{L-M}\big(\specbaz -\tilde{\specbaz}_k(\specsubtraction)\big),
\end{equation}
where $\tilde{\specbaz}_0(\specsubtraction)=\specsubtraction$. The eigenvalues of all the other states follow from $\alg{su}(2)$ invariance.

\subsection{Numerical results on the root distributions of \texorpdfstring{$Q(u)$}{} and \texorpdfstring{$P(u)$}{}}
\label{sec:numerics}

In the previous sections\footnote{Recall that for the numerical studies of this section we will return from the $\specbaz$-plane to the Wick-rotated $\specfad$-plane via \eqref{spectralparameters}.} we obtained for twist $\twist=0$ a one-parameter family of a {\it pair} of operators $\{\Qop(\specfad),\Pop(\specfad)\}$. On the level of eigenvalues, this free complex parameter changes the overall normalization of the polynomial eigenvalues $Q(\specfad)$ in \eqref{baxterpolyPQ}, and both the normalization and the position of the roots of the polynomial eigenvalues $P(\specfad)$ in \eqref{baxterpolyPQ}. Let us numerically study the positions of the roots of these two complementary polynomials for various interesting classes of states. Here clearly the overall normalization does not matter, and we can use the parameter to shift one of the $L-M+1$ roots of $P(\specfad)$ to any position in the complex $\specfad$ plane we fancy.

We will focus here only on the case where both $L$ and $M$ are even. Let us start with the polynomial $Q(u)$. From the definition it is a polynomial of degree $M$ with roots being the solutions of Bethe equations (\ref{betheeqsnotwist}). With every Bethe root $u_k$ we identify an integer $n_k$, the so-called mode number, which labels different solutions. Every solution is then characterized by its set of mode numbers. To keep the discussion as transparent as possible we will focus only on specific choices of this set. Namely, we will consider two-cut solutions by taking symmetric distribution of roots on the complex plane with all mode numbers equal $|n_k|=n$ for $k=1,\ldots,M$. The method of finding such solutions can be found in \cite{Bargheer:2008kj} (see Appendix \ref{app:numerics} for details). With the algorithm given there we are however only able to find solutions of (\ref{betheeqsnotwist}) when restricting to $\frac{M}{L}<\frac{1}{2}$, i.e.~below half-filling. In order to then also find the dual roots from (\ref{dualbetheeqsnotwist}), which corresponds to the polynomial $P(u)$, we have to find a different method. The idea is to use the generalized Wronskian relation (\ref{TopjPQfinal}) with $\phi=0$ and $j=0$. Then the left hand side of (\ref{TopjPQfinal}) is exactly $u^L$, and the problem of finding the polynomial $P(u)$ reduces to a linear problem for its $L-M+2$ coefficients. Having the polynomial, finding its roots is a trivial numerical task. 

\begin{figure}[t]
\begin{center}
\includegraphics{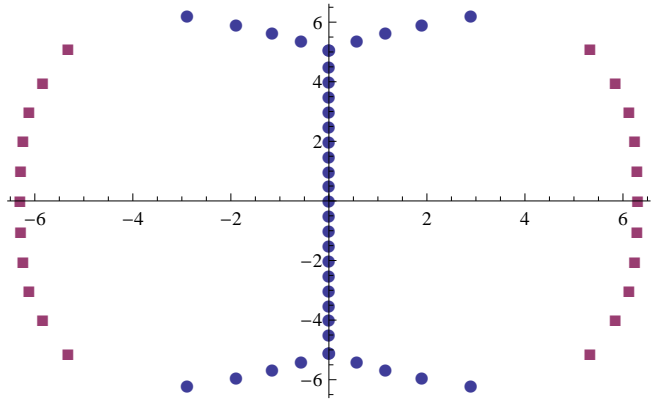}
\end{center}
\caption{The two-cut root distribution of the polynomials $Q(u)$ (purple squares) and $P(u)$ (blue dots). Here $L=60$, $M=22$ and $|n_k|=1$ for all $k$.}
\label{L60M22}
\end{figure}

Using the methods presented above, we were able to find many different configurations. One typical example is given in figure \ref{L60M22}, see also \cite{Beisert:2003xu} and \cite{Gromov:2007ky}. Here the length is $L=60$ and the number of Q-roots is $M=22$. As we mentioned before, there is a freedom in choosing the position of one of the roots of the polynomial $P(\specfad)$. We took it to be $\tilde{u}_0=0$\footnote{In the Appendix \ref{app:numerics} we will also show results for other choices.}. As was pointed out in \cite{Bargheer:2008kj}, the positions of the roots of the polynomial $Q(\specfad)$  are well described by the predictions from the thermodynamic limit -- they lie almost on a ``spectral curve''. On the other hand, the roots of the polynomial $P(\specfad)$ are located on a dual configuration of cuts (see fig. 40 in \cite{Bargheer:2008kj}). It means that $Q$ and $P$ establish two different ways of slicing up the complex plane, with cuts originating from the branch points given by thermodynamic calculations.

We will end our discussion of Q- and P-roots with a few comments on the {\it mode numbers} $n_k$ relevant to the above examples. In order to define these, the untwisted Bethe equations \eqref{betheeqsnotwist} need to be rewritten in logarithmic form
\begin{equation}\label{BE.logaritmic.form}
L\, \log \frac{\specfad_k+\frac{i}{2}}{\specfad_k-\frac{i}{2}} +2\,\pi\, i\, n_k=\sum_{j\neq k}\, \log\frac{\specfad_k-\specfad_j+i}{\specfad_k-\specfad_j-i}\, ,
\end{equation}
where the summation is over all roots of either the Q- or the
P-polynomial. We can immediately check that for the Q-roots presented
in fig.~\ref{L60M22} the mode numbers are $+1$ for the roots with
positive real part and $-1$ for the ones with negative real part. The
same is true for the P-roots, but in this case there are also other
roots which are purely imaginary and form a so-called
condensate, which is nothing but a giant string in the standard
terminology of the Bethe Ansatz (to be precise, in the case of
fig.~\ref{L60M22} it is a combination of two strings shifted by $\frac{i}{2}$).
As explained at the end of section~\ref{sec:bethe} the strings are {\em
  not exact}, however their roots very closely approach the (half)
integer positons on the imaginary axis (the roots at
$u=\pm\frac{i}{2}$ are exact). 
The mode numbers for the roots in the strings, are not well defined 
due to the ambiguity of defining logarithms of real negative
arguments, appearing in r.h.s. of \eqref{BE.logaritmic.form}.


\section{Relation to Earlier Work}
\label{sec:earlierwork}

\subsection{Historical note.}

The concept of the $\Qop$-operator had been introduced by Baxter in
his seminal paper \cite{Baxter:1972hz} on the symmetric 8-vertex
model and further developed in this context in \cite{Baxter:1972wf,Baxter:book} 
and more recently in \cite{Fabricius:2002bi,Bazhanov:2004hv}.
Baxter's idea 
received a major boost in \cite{Bazhanov:1996dr}, in an attempt to
understand the integrable structure of conformal field theory, which
is related to the continuum limit of the integrable XXZ model.
Renewed attention to the $\Qop$-operator arose independently in the
context of the discovery of non-compact integrable Heisenberg magnets
in the high energy scattering problem of QCD
\cite{Lipatov,Faddeev:1994zg}.
The emergence of integrability in the spectral problem of planar $\mathcal N=4$ and in AdS/CFT \cite{Lipatov:1997vu,Braun:1998id,Minahan:2002ve,Bena:2003wd}, not yet understood beyond the one-loop level, is the motivation behind our fresh look at the $\Qop$-operator in this article.

In the present chapter we will very briefly review what was previously  known about the $\Qop$-operator for the XXX spin $j$ Heisenberg magnet in order to compare with our findings above. In the comparison between different approaches to the problem it seems to be important to distinguish between 
compact magnets ($2j \in \mathbb{Z}_{\geq 0}$) and non-compact magnets, where typically either $2j=-1,-2,\ldots$ (discrete series representations of $\alg{su}(1,1)$ or $\alg{sl}(2,\mathbb{R})$), or else $j$ takes on certain continuous values (principal and supplementary series representations). The quantum space is indeed very different in the compact and non-compact cases, as it is finite dimensional in the former, and infinite dimensional in the latter for any length $L$ of the chain.  


\subsection{Non-compact magnets}
\label{sec:earliernoncomp}
 In the case of non-compact magnets 
two main ways to built Baxter's $\Qop$-operator(s) have been proposed.
These are related, but this fact has not yet been properly investigated. 
The first way is based on the connection between B\"acklund transformations in the theory of classical 
integrable systems and Baxter's $\Qop$-operator for quantum integrable systems.  Sklyanin et al.~(see \cite{Sklyanin:2000} and references therein),
inspired by \cite{Gaudin:1992ci}, proposed a way to build the $\Qop$-operator as an integral operator for models governed by the $\alg{sl}(2)$ invariant R-matrix.
It is constructed as the trace over some monodromy built from Lax
operator intertwining operators connected with the Toda, DST (discrete self-trapping) and XXX models, respectively. An explicit construction for the XXX model has been proposed in \cite{Choudhury:2003}. 

The second way is based on some factorization properties of 
{\em matrix elements} (related but different to operator factorization
of this paper) of 
$R$-matrices and $L$-operators, which, apparently, was first employed
in \cite{Bazhanov:1989nc} in the context of the chiral Potts model \cite{vonGehlen:1984bi,AuYang:1987zc,Baxter:1987eq}.
This approach allows to explicitly calculate {\em matrix elements} of
the ${\bf Q}$-operator as a simple product involving only two-spin
factors. The same method was later used in \cite{Gaudin:1992ci} for the
Toda model then in a series of papers \cite{Derkachov:2005xn,Derkachov:1999} 
studying various aspects of the factorization of $R$-matrices,
associated with the XXX model.  It would be interesting to clarify 
the exact relationship (if any) between the factorization in these works and the one used in this paper.


\subsection{Compact magnets}
\label{sec:earliercomp}

Among the compact magnets XXX$_{1/2}$ is the most studied, beginning with Bethe's original work \cite{Bethe}. Here we will restrict to this case. While a lot is known about the $Q$-functions, i.e. the eigenvalues of the $\Qop$-operator(s), much less is known about the construction of the  latter as a trace over a suitable monodromy matrix for  XXX$_{1/2}$.

In \cite{Pronko:1998xa}, inspired by \cite{Bazhanov:1998dq},  Baxter's equation, at the level of eigenvalues, had been considered.
For a given eigenvalue of the $\Top$-operator acting on some h.w.s, two linearly independent polynomial solutions
of Baxter's equation were found. These two solutions were called $P$ and $Q$ in \cite{Pronko:1998xa}, and are generically distinguished by their degree of polynomiality in the spectral parameter. It was furthermore shown that the lower degree polynomial $Q$ can always be added to the higher degree polynomial $P$, leading to a one parameter freedom in the root distribution of the higher degree polynomial.

Subsequently Pronko built a one-parameter family of $\Qop$-operators in \cite{Pronko:1999is}, which we denoted in section \ref{sec:Popzerotwist} $\Pop$-operators. However, as explicitly written in equation \eqref{eq:oneparameterPfromT}, the one-parameter family of $\Pop$-operators is nothing but the set of all $\Top$-operators after a linear redefinition of their spectral parameter and spin.
Thus the construction in  \cite{Pronko:1999is}, following \cite{Baxter:1972hz}, finally boils down to
the construction of  $\Top_j(\specfad)$ for general, complex values of the spin $j \in \mathbb{C}$.
As discussed in \Appref{app:finiteTj} there are various seemingly different but ultimately equivalent procedures to achieve this.

In \cite{Pronko:2000} and \cite{kovalsky} Pronko tried to build two ``basic'' $\Qop^\pm$ instead of a one parameter family of $\Qop$-operators. He succeeded in doing this for the Toda chain and the DST ``discrete self trapping'' chain, respectively.
Still it is important to keep in mind that his construction, apart from the fact that it was applied to a different class of models, remains conceptually different from the one proposed in this work.
The idea in \cite{Pronko:1999is,Pronko:2000,kovalsky} is to build some operator satisfying Baxter's equation.
In contradistinction, in the present paper  we aim at the fundamental relation \eqref{fundrel}. The factorization formulas 
\eqref{firstway-final}, \eqref{secondway-final} are the crucial step in doing this.
The relation between the two constructions deserves further investigations.

\section{Conclusions and Open Problems}
\label{sec:openproblems}

In this work we have demonstrated that the complete understanding of the ``simplest'' of all integrable quantum models, the {\it compact} spin-$\half$ XXX Heisenberg model, requires the introduction of {\it non-compact} oscillator representations. After this is done, a very interesting factorization of the compact quantum Lax-operator takes place, cf.~\eqref{firstway-final}, \eqref{secondway-final}. This fact may then be used to quickly find a very elegant solution of the model in terms of the full set of {\it operatorial} (as opposed to mere equations for their eigenvalues) functional relations between a large class of commuting families of transfer matrices, which naturally includes two linearly independent Baxter $\Qop$-operators. An added benefit is that all transfer matrices, including the Baxter $\Qop$-operators, are explicitly constructed as traces over suitable monodromy matrices. The latter is particularly important, as the {\it analytic structure} of the solutions to the hierarchy does not have to be guessed, but is clear from the construction. In this sense we consider our approach a first sketch of what needs to be done in order to understand the solution of the AdS/CFT spectral problem proposed in \cite{Arutyunov:2009zu}, where the analytic structure of the solutions of the infinite set of functional integral equations obtained from the TBA approach has to be guessed and then imposed by hand. 

Apart from this technical advantage of a direct construction on the operatorial level, we believe that there will eventually be an interesting physical interpretation of our procedure. Baxter $\Qop$-operators enter the T-system of AdS/CFT as ``boundary values''. Furthermore, it is clear that the auxiliary channel in which the bosonic excitations we need for our construction are propagating is closely related to the ``mirror direction'' in the TBA approach. A similar analogy was recently pointed out in \cite{Gromov:2010vb}. We believe that these kinds of fundamental short-range excitations are suitable for replacing the long-range ``mirror magnons'' of AdS/CFT.

An important feature of our approach is that a magnetic flux (horizontal field, or twist) has to be applied in order to ensure convergence of the Baxter $\Qop$-operators. This suggests that in some sense the theory with the flux is more natural, and the untwisted case is a somewhat singular limit. It is possible that this feature also lifts to the AdS/CFT case. As was stated in the introduction, in AdS/CFT the twisted theory also appears to be integrable \cite{Beisert:2005if}, but much less is known about it. In particular, its finite-size structure is very puzzling, see e.g.~\cite{Bykov:2008bj}. In any case, from our point of view it is still not clear whether one can construct Baxter $\Qop$-operators in the absence of horizontal fields directly as a trace over some monodromy matrix. 

Clearly our approach should be extended to the compact $\alg{sl}(n)$ and the supersymmetric $\alg{sl}(n|m)$ cases, as well as the non-compact magnets, with the goal of treating the symmetry algebra $\alg{su}(2,2|4)$ of AdS/CFT \cite{blms}. In particular it will be interesting to relate our method to the Baxter $\Qop$-operator construction developed for non-compact spin chains in \cite{Derkachov:2005xn,Derkachov:1999}.

Finally we would like to mention yet another interesting research direction in the theory of
${\bf Q}$-operators. It concerns their connection to the spectral theory
of differential equations \cite{Voros,Dorey:1998pt,Bazhanov:1998wj}.
It recently found new applications in the theory of BPS states of a large class of $d=4$, ${\cal N}=2$ field theories \cite{Gaiotto:2009hg}, as well as in the (mysteriously) closely related theory of classical string solutions for certain strongly coupled Wilson loops in ${\cal N}=4$ gauge theory \cite{Alday:2009dv}. Lastly this spectral theory also appeared in the description of the
massive Sine(h)-Gordon model \cite{Lukyanov:2010rn}. For the XXX model
the connection to differential equations arises in the $c=1$ CFT
limit \cite{Bazhanov:2003ua}, which is closely related to the Kondo model \cite{Andrei:1982cr}.


\subsection*{Acknowledgments}
We would like to thank Changrim Ahn, Till Bargheer, Niklas Beisert, Sergey Frolov,
Wellington Galleas, Alexander Molev, Tetsuji Miwa, Michio Jimbo, Vladimir Kazakov, Gregory Korchemsky, Vladimir Manga\-zeev, Fedor Smirnov and Pasha Wiegmann for useful discussions. Special thanks to Till Bargheer for his help on the numerical work, and to Stefan Zieme for a careful reading of the manuscript. T.~{\L}ukowski was supported by Polish science funds during 2009-2011 as a research project (NN202 105136). 
He also acknowledges the support of the DAAD in form of an ``Auslandsstipendium'' which allowed him an extended visit at the AEI Potsdam during which this project was worked out. M.~Staudacher thanks the Department of Theoretical Physics at ANU, Canberra, as well as the IEU, Seoul for hospitality while working on this project. 

\appendix

\section{Functional relations}
\label{app:functionalrel}

\noindent
Using  the defining relations for $\mathcal T$ and $\Top$ in \eqref{fundrelcompact} 
and \eqref{fundrel}, one finds
\begin{equation}
 \mathcal{T}_{AB} = f(\twist) \, \Top_{\sfrac{A-B-1}{2}}(\sfrac{A+B}{2})\, .
\end{equation}
A more explicit form of the left equation in \eqref{funcrelcompact} reads
\begin{equation}
\mathcal{T}_{CB}\,\Qop^{\pm}_A
=
\mathcal{T}_{AB}\,\Qop^{\pm}_C+\mathcal{T}_{CA}\,\Qop^{\pm}_B
\, .
\end{equation}
Choosing
\begin{equation}
 A=\specbaz\,, \qquad B=\specbaz+\specbaz'+j +\half\,, \qquad C=\specbaz+\specbaz'-j -\half\,,
\end{equation}
one gets
\begin{eqnarray}
 \Top_{j}(\specbaz+\specbaz')\,\Qop_{\pm}(\specbaz)
&=&
 \Top_{\half(j-\specbaz' -\half)} \left(\specbaz + \half(\specbaz' - j -\half)\right)\,
\Qop_{\pm} \left(\specbaz+\specbaz'+j +\half \right)+\nonumber \\
& &
 \Top_{\half(j+\specbaz' -\half)} \left(\specbaz + \half(\specbaz' + j +\half)\right)\,
\Qop_{\pm} \left(\specbaz+\specbaz'-j -\half \right).
\end{eqnarray}
Analogously the right equation in \eqref{funcrelcompact} reads
\begin{equation}
\mathcal{T}_{CB}\mathcal{T}_{AD}
=
\mathcal{T}_{AB}\,\mathcal{T}_{CD}+\mathcal{T}_{CA}\,\mathcal{T}_{BD}
 \, .
\end{equation}
With the choice
\begin{equation}
 A=\specbaz' +j'+\half\,, \qquad D=\specbaz' -j'-\half\,\,, \qquad C=\specbaz+j +\half\,\qquad B=\specbaz-j -\half\,,
\end{equation}
one gets
\begin{eqnarray}
\Top_{j}(\specbaz)\,\Top_{j'}(\specbaz')
&=&
\Top_{\sfrac{\specbaz-\specbaz' +j-j'-1}{2}} \left(\sfrac{\specbaz+\specbaz'+j+j'+1}{2}\right)\,
\Top_{\sfrac{\specbaz-\specbaz' -j+j'-1}{2}} \left(\sfrac{\specbaz+\specbaz' -j-j'-1}{2}\right) 
+\nonumber \\
& &
\Top_{\sfrac{\specbaz'-\specbaz +j'+j}{2}} \left(\sfrac{\specbaz'+\specbaz +j'-j}{2}\right)\,
\Top_{\sfrac{\specbaz-\specbaz' +j+j'}{2}} \left(\sfrac{\specbaz+\specbaz' +j-j'}{2}\right).
\end{eqnarray}
%

\section{Explicit results for small chain lengths}
\label{app:smalllengths}

In this appendix, where we revert to the notation\footnote{%
The reason is that the Bethe equations are most commonly written in the $\specfad$-plane.}
with the spectral parameter $\specfad=i \specbaz$, cf.~\eqref{spectralparameters}, we will present the results for $\Qop_-(\specfad)$, $\Qop_+(\specfad)$ and  $\Top(\specfad)$ for the smallest possible chain lengths\footnote{
Note that while it is not possible to define a nearest neighbor Hamiltonian for a $L=1$ ``chain'', the definitions \eqref{monodromy},\eqref{transfertrace} for the transfer matrix and \eqref{Qpm} for the Baxter operators still make sense.}
$L=1,2$. All matrix elements of $\Qop_-(\specfad)$, $\Qop_+(\specfad)$ are easily computed from \eqref{Qpm}. For the convenience of the reader we will present the ``Wick-rotated''  (i.e. written in the $\specfad$-plane) Lax operators for the Baxter $\Qop$-operators. Recall also the definition $z_\pm=z\pm(j+\half)$ in \eqref{lightconespec}.  They read, with $\osch=\osca^+\,\osca^-+\half$ as defined in \eqref{harm},
\beq\label{Lplusminusu}
\Lbaz^-(\specfad)=
\left( \begin{array}{cc}
1 \ \  & ~\osca^+\, \, \\
i\,\osca^- \ \  \,&\specfad+i\,\osch
\end{array} \right)\ .
\eeq
and
\beq\label{Lplusu}
\Lbaz^+(\specfad)=
\left( \begin{array}{cc}
\specfad-i\,\osch \, \, & ~~i\,\osca^+\, \, \\
- \osca^- \, \,&~~1\, \, \end{array} \right)\ .
\eeq
The definitions of $\Lbaz^\pm$ given above differ from the definitions $L_\pm$
given in  \eqref{Lplus}, \eqref{Lminus} in the main text. The two definitions are related in this way
\beq \label{Lrelations}
\Lbaz^-(i \specbaz)=
\left( \begin{array}{cc}
1\, & ~0\, \, \\
0 \, & ~i \, \end{array} \right)\,L_-(\specbaz)\,, \qquad
\Lbaz^+(i \specbaz)=
\left( \begin{array}{cc}
i\, & ~0\, \, \\
0 \, & ~1 \, \end{array} \right)\,L_+(\specbaz)\,.
\eeq
The conventions used in this appendix are such that the eigenvalues of $\Qop_\pm$ have the form
$\exp{(\pm \sfrac{\twist}{2} u)} (u^p+ \dots)$, where $p$ is some integer.
Notice that the position of the diagonal matrix (left or right) in \eqref{Lrelations} can be compensated by an oscillator algebra automorphism.
In particular this means that left and right multiplication became equivalent after the trace is taken. This in nothing but restating the U$(1)$ invariance of $\Qop_{\pm}$. The Baxter operators \eqref{Qpm} become in the $\specfad$-notation
\begin{equation}
\label{Baxteru}
\Qop_\pm(u;\twist) \equiv \frac{e^{\pm \frac{\phi}{2} \specfad}}{\mbox{Tr}_{\mathcal F}(e^{- i\, \twist\, h})}
\mbox{Tr}_{\mathcal F} \left(e^{-i \,\twist\, h }\, \Lbaz^\pm_L (\specfad) \otimes \cdots \otimes \Lbaz^\pm_1 (\specfad)\right)\, .
\end{equation}
%
%
%
%
The matrix elements of the transfer matrix $\Top(\specfad)$ are then obtained, for $j=\half$, from \eqref{fundrel}. It is easy to check that the direct construction based on \eqref{monodromy}, \eqref{transfertrace} leads to the same results.
We will also write down the $L=1,2$ eigenvalues $Q_-(\specfad)$, $Q_+(\specfad)$ and  $T(\specfad)$ of these operators.

It is easy to check that the operators below indeed satisfy the $j=0$ and $j=\half$ cases of the Wronskian relation \eqref{fundrel} translated to $u$-space
\begin{eqnarray}
2\,i\sin\frac{\phi}{2}\, u^L
&=&
\Qop_+(\specfad+\frac{i}{2})\,\Qop_-(\specfad-\frac{i}{2})-\Qop_+(\specfad-\frac{i}{2})\,\Qop_-(\specfad+\frac{i}{2}),\\
2\,i\sin\frac{\phi}{2}\, \Top(\specfad)
&=&
\Qop_+(\specfad+i)\, \Qop_-(\specfad-i)-\Qop_+(\specfad-i)\,\Qop_-(\specfad+i).
\end{eqnarray}
%


\subsection{Length \texorpdfstring{$L=1$}{}}
 
\subsubsection{\texorpdfstring{$\Qop_-(\specfad)$}{}}
\begin{equation*}
e^{-\frac{\phi}{2}u}\left(
\begin{array}{cc}
 1 & 0 \\
 0 & \,\, u+\frac{1}{2} \cot \frac{\phi }{2}
\end{array}
\right)
\end{equation*} 

\begin{equation*}
e^{-\frac{\phi}{2}u}\left\{1\, ,\, u+\frac{1}{2} \cot \frac{\phi }{2}\right\}
\end{equation*}

\subsubsection{\texorpdfstring{$\Qop_+(\specfad)$}{}}
\begin{equation*}
e^{\frac{\phi}{2}u}\left(
\begin{array}{cc}
 u-\frac{1}{2} \cot \frac{\phi }{2} & \,\,0 \\
 0 & \,\,1
\end{array}
\right)
\end{equation*}

\begin{equation*}
e^{\frac{\phi}{2}u}\left\{1,u-\frac{1}{2} \cot \frac{\phi }{2}\right\}
\end{equation*}

\subsubsection{\texorpdfstring{$\Top(\specfad)$}{}}
\begin{equation*}
\left(
\begin{array}{cc}
 2 u \cos \frac{\phi }{2}-\sin \frac{\phi
   }{2} & 0 \\
 0 & 2 u \cos \frac{\phi }{2}+\sin \frac{\phi
   }{2}
\end{array}
\right)
\end{equation*}

\begin{equation*}
\left\{2 u \cos \frac{\phi }{2}-\sin \frac{\phi
   }{2},2 u \cos \frac{\phi }{2}+\sin
   \frac{\phi }{2}\right\}
\end{equation*}


\subsection{Length \texorpdfstring{$L=2$}{}}

\subsubsection{\texorpdfstring{$\Qop_-(\specfad)$}{}}
\begin{equation*}
e^{-\frac{\phi}{2}u}\left(
\begin{array}{cccc}
 1 & 0 & 0 & 0 \\
 0 &\, \, u+\frac{1}{2} \cot \frac{\phi }{2}\,\, &\,\, \frac{1}{2} \cot \frac{\phi }{2}+\frac{i}{2} & 0 \\
 0 & \frac{1}{2} \cot \frac{\phi }{2}-\frac{i}{2} \,\,& \,\,u+\frac{1}{2} \cot \frac{\phi }{2} & 0 \\
 0 & 0 & 0 & u^2+u\cot \frac{\phi }{2} +\frac{1}{2\sin^2 \frac{\phi}{2}}-\frac{1}{4}
\end{array}
\right)
\end{equation*}

\begin{equation*}
e^{-\frac{\phi}{2}u}\left\{1\, ,\, u+\frac{1}{2} \cot \frac{\phi }{4}\, ,\, u-\frac{1}{2} \tan \frac{\phi }{4}\, ,\, u^2+u \cot \frac{\phi
   }{2}+\frac{1}{2\sin^2 \frac{\phi}{2}}-\frac{1}{4}\right\}
\end{equation*}

\subsubsection{\texorpdfstring{$\Qop_+(\specfad)$}{}}
\begin{equation*}
e^{\frac{\phi}{2}u}\left(
\begin{array}{cccc}
 u^2-\cot \frac{\phi }{2} u+\frac{1}{2\sin^2 \frac{\phi}{2}}-\frac{1}{4}\,\, & 0 & 0 & 0 \\
 0 & u-\frac{1}{2} \cot \frac{\phi }{2} \,\,&\,\, -\frac{1}{2} \cot \frac{\phi }{2}-\frac{i}{2}\,\, & 0 \\
 0 & -\frac{1}{2} \cot \frac{\phi }{2}+\frac{i}{2} \,\,&\,\, u-\frac{1}{2} \cot \frac{\phi }{2} & 0 \\
 0 & 0 & 0 & 1
\end{array}
\right)
\end{equation*}
\begin{equation*}
e^{\frac{\phi}{2}u}\left\{1\, ,\, u-\frac{1}{2} \cot \frac{\phi }{4}\, ,\, u+\frac{1}{2} \tan \frac{\phi }{4}\, ,\, u^2-u \cot \frac{\phi
   }{2}+\frac{1}{2\sin^2 \frac{\phi}{2}}-\frac{1}{4}\right\}
\end{equation*}

\subsubsection{\texorpdfstring{$\Top(\specfad)$}{}}
\begin{equation*}
\left(
\begin{array}{cccc}
 \left(2 u^2-\frac{1}{2}\right) \cos \frac{\phi }{2}-2 u
   \sin \frac{\phi }{2} & 0 & 0 & 0 \\
 0 &  \left(2 u^2+\frac{1}{2}\right) \cos \frac{\phi
   }{2} & -e^{\frac{i \phi }{2}} & 0 \\
 0 & -e^{-\frac{i \phi }{2}} & \left(2 u^2+\frac{1}{2}\right) \cos
   \frac{\phi }{2} & 0 \\
 0 & 0 & 0 & \left(2 u^2-\frac{1}{2}\right) \cos \frac{\phi
   }{2}+2 u \sin \frac{\phi }{2}
\end{array}
\right)
\end{equation*}
\begin{eqnarray*}
&&\left\{\left(2 u^2-\frac{1}{2}\right) \cos \frac{\phi
   }{2}-2 u \sin \frac{\phi }{2},\left(2 u^2+\frac{1}{2}\right) \cos \frac{\phi
   }{2}+1,\right. \\
&& \left. \left(2 u^2+\frac{1}{2}\right) \cos \frac{\phi
   }{2}-1, \left(2
   u^2-\frac{1}{2}\right) \cos \frac{\phi }{2}+2 u \sin
   \frac{\phi }{2}\right\}
\end{eqnarray*}


\section{Analytic continuation of the trace}
\label{app:finiteTj}

In this appendix we will review three ways of defining traces for complex spin. The first way, used in our construction in the main body of the paper, involves introducing a twist $\twist$ as a regulator and then takes $\twist \rightarrow 0$ while subtracting infinities. The second way has been introduced in \cite{Boos:2004zq} under the name ``trace functional''.
The third way has been used by Pronko in \cite{Pronko:1999is}. We will now show the equivalence of these three constructions. In the case of the first two, this had already been noticed in \cite{Korff:2006}.

Let us first fix the set-up for the analysis. Denote by $\Op$ any one of the $2^L \times 2^L$ matrix elements of the monodromy matrix,
see  \eqref{genmonodromy}. $\Op$ is some expression written in terms of the $\alg{su}(2)$ generators 
$\SpinAux^0,\SpinAux^{\pm}$ acting in the auxiliary space
(we will also use the notation $\SpinAux_k^l$).
It is clear from $\alg{u}(1)$ invariance that any such $\Op$ has a definite grading under $\SpinAux^0$, namely:
\begin{equation}
\label{zeroweight}
 [\SpinAux^0 ,\Op]= s(\Op)\, \Op\,.
\end{equation}
If $s(\Op) \ne 0$, then its trace will vanish for any complex value of the spin.
So let us consider the case $s(\Op) = 0$. In this case  $\Op$ can be rewritten as 
\begin{equation}
\label{opzeroweight}
\Op \rightarrow F(\SpinAux^0, \vec \SpinAux^2),
\end{equation}
where $F$ is some function.


\subsection{Twist regularization and \texorpdfstring{$\twist \rightarrow 0$}{}  limit}

Let us consider
\begin{equation}
\label{eq:FinitenessTj}
 \Tr_{\pi_j^+}\left( e^{-i \,\twist \,\SpinAux^0} \Op \right)- \Tr_{\pi_{-j-1}^+}\left( e^{-i \, \twist \, \SpinAux^0} \Op \right),
\end{equation}
where the trace is taken over the infinite dimensional Verma module.
We will show that this quantity is finite in the $\twist \rightarrow 0$ limit.

To proceed it is useful to realize the highest weight representation explicitly
in a Fock space. The latter is an infinite module $\pi_l^+$ for any $l$. It thus contains the information about the spin in our realizations of the generators, see \eqref{HPfirst} and \eqref{HPsecond}. 
As just discussed, we will take $\Op$ as in \eqref{opzeroweight}. Then (\ref{eq:FinitenessTj}) becomes
\begin{equation}
\label{FinitnessinFock} 
\sum_{n=0}^\infty \left( e^{-i \twist (n - j)} F(n - j,  j(j+1))- e^{-i \twist (n +j +1)} F(n +j+1,  j(j+1)) \right).
\end{equation}
Introducing $f(n-\nu )\equiv F(n-j,j(j+1))$, where $\nu= j+ \frac{1}{2}$,  and we suppress the dependence on the Casimir $j(j+1)$, \eqref{FinitnessinFock} can be rewritten as
\begin{equation}
\label{Finitnessnu}
e^{-i\frac{\twist}{2}}\sum_{n=0}^\infty \left( e^{-i \twist (n - \nu)} f(n-\nu )-  e^{-i \twist (n +\nu)} f(n+\nu ) \right).
\end{equation}
Consider the case $f(s)= s^k$ for some integer $k$. Then  (\ref{Finitnessnu}) reads
\begin{equation}
e^{-i\frac{\twist}{2}} \left(i \frac{\partial}{\partial \twist} \right)^k
 \sum_{n=0}^\infty \left( e^{-i \twist (n - \nu)}-  e^{-i \twist (n +\nu)}\right)
 =
 e^{-i\frac{\twist}{2}} \left(i \frac{\partial}{\partial \twist} \right)^k \left( e^{i\frac{\twist}{2}}
 \frac{\sin \nu \twist }{\sin \frac{\twist}{2}} \right).
\end{equation}
This expression is finite in the $\twist \rightarrow 0$ limit for any $k, \nu$.
This concludes the proof of the finiteness of $\Top_j(\specbaz;0)$


\subsection{Trace functional}

We now want to show that the procedure of evaluating traces over complex spin
of the previous section is the same as the one used in \cite{Boos:2004zq} and termed ``trace functional''. 
It has the property that\footnote{
For convenience of the reader the notation used
here is related to the one used in  \cite{Boos:2004zq} in the following way: $x=2\, \nu$, $H=2 \, \SpinAux_0$, $z=\frac{t}{2}$.
See \cite{Boos:2004zq} for the complete set of defining properties of \eqref{tracefunctional}.}
\begin{equation}\label{tracefunctional}
 \Tr^{\text{functional}}_{\nu} e^{t \, \SpinAux_0}\equiv\,\frac{\sinh(t\, \nu)}{\sinh{\sfrac{t}{2}}}  \,. 
\end{equation}
This formula plays the role of a generating function for the traces of $(\SpinAux_0)^k$.
To show the equivalence with the definition of the previous section it is enough to
consider \eqref{eq:FinitenessTj}  with  $\mathcal{O}=1$ and $-i \twist =t$. This gives
\begin{equation}
\frac{\sinh(t \, \nu)}{\sinh{\sfrac{t}{2}}}
\, , 
\end{equation}
and completes the proof of equivalence.


\subsection{Other approaches}

We will now show that there is a third way to define traces over complex spin, used in  \cite{kovalsky}. It is equivalent to the previous two procedures. Now the definition of the trace is 
\begin{equation}
\Tr^{\text{Pronko}}_{\nu}\, \Op(\SpinAux_k^l)\equiv \int d \mu (\gamma,\bar \gamma) \, \Op(\gamma^l \frac{\partial}{\partial \gamma^k})
 \, \frac{\left(\gamma \cdot \bar \gamma\right)^{2 \nu -1}}{\Gamma(2 \nu )} \,, 
\end{equation}
where
\begin{equation}
 d \mu (\gamma,\bar \gamma) \equiv e^{-\gamma \cdot \bar \gamma}\, \prod_{k=1}^2 \,\frac{d \gamma^k d \bar \gamma_k} {2 \pi i } \,,
\qquad
 \gamma \cdot \bar \gamma \equiv \gamma^1 \bar \gamma_1 +\gamma^2\bar \gamma_2 \,. 
\end{equation}
As discussed before only the zero weight part of $\Op(\SpinAux_k^l)$ will contribute to the trace, cf.~\eqref{zeroweight}.
To make contact with the previous analysis we compute
\begin{align}
\Tr^{\text{Pronko}}_{\nu}\,e^{t\,  \SpinAux_0} & = \int d \mu (\gamma,\bar \gamma) \,
 e^{\sfrac{t}{2}\left(\gamma^1 \frac{\partial}{\partial \gamma^1}-\gamma^2 \frac{\partial}{\partial \gamma^2} \right)}
\, \frac{\left(\gamma \cdot \bar \gamma\right)^{2 \nu -1}}{\Gamma(2 \nu )} \nonumber \\ 
& = \frac{1}{\Gamma(2 \nu )}\int d \mu (\gamma,\bar \gamma) \, 
\left( e^{\sfrac{t}{2}}\,\gamma^1 \,\bar \gamma_1+e^{\sfrac{t}{2}}\,\gamma^2\,\bar \gamma_2\right)^{2 \nu -1} \nonumber \\ 
& =\frac{\sinh(t \, \nu)}{\sinh{\sfrac{t}{2}}}\,.
\end{align}
This last equation shows the equivalence with the other two definitions of trace. 

\section{\texorpdfstring{$\alg{su}(2)$}{} covariance properties}
\label{app:sl2inv}

In this appendix we will study some $\alg{su}(2)$ transformation properties of $\Qop_\pm$.
Thanks to the identity
\begin{equation}
e^{\gamma\, \SpinQ^-} 
\left( \begin{array}{cc}
1 \,& \osca^+\, \\
\osca^-\, & \specL+\,\osca^+\,\osca^-\,\end{array} \right)
=
e^{-\gamma \,\osca^+} 
\left( \begin{array}{cc}
1 \,& \osca^+ \,\\
\osca^- \,& \specL+\osca^+\,\osca^-\end{array}\, \right)
e^{\gamma \,\osca^+}\,,
\end{equation}
where
\begin{equation}
e^{\gamma\, \SpinQ^-}
=\left( \begin{array}{cc}
1\,\,& 0 \,\,\\
\gamma \,\,& 1\,\, \end{array} \right)\,,
\end{equation}
the trace of a monodromy matrix built from 
\begin{equation}
\left( \begin{array}{cc}
1 \, & \osca^+\, \\
\osca^- \,& \specL+\osca^+\,\osca^-\,\end{array} \right)\,,
\end{equation}
 will satisfy the equation
 \begin{equation}
e^{\gamma \, \SpinQ^-_{\text{tot}}}\, \Qop_-(\specL)=\Qop_-(\specL)\,.
\end{equation}
This is a formal equation because for the trace to give a finite result one needs to  
 introduce a regulator $e^{- i \, \twist \, \osca^+ \, \osca^-}$, and this operator  does not commute with $e^{\pm \, \gamma \, a^+}$. Still one can write
 \begin{equation}
e^{\gamma \, \SpinQ^-_{\text{tot}}}\,\Qop_-(\specL; \twist)=
\Qop_-(\specL;\twist) \, \left( \Identity + \mathcal O (\twist)\right)\,.
\end{equation}
In a similar way one can show that
 \begin{equation}
\Qop_+(\specLbar;\twist)\, e^{\gamma \,  \SpinQ^+_{\text{tot}}}=
\Qop_+(\specLbar;\twist) \, \left( \Identity + \mathcal O (\twist)\right)\,.
\end{equation}
On the other hand $\Qop_\pm$ are $\alg{u}(1)$ invariant for any value of the twist, namely
 \begin{equation}
e^{\gamma \, \SpinQ^3_{\text{tot}}} \, \Qop_{\pm}(\specbaz_{\pm};\twist)\, e^{-\gamma \, \SpinQ^3_{\text{tot}}}=
\Qop_{\pm}(\specbaz_{\pm};\twist) \,.
\end{equation}
Using the previous identities one can show that
\begin{equation}
e^{\vec \gamma \, \vec  \SpinQ_{\text{tot}}}\, \Qop_{\pm}(\specbaz^1_{\pm};\twist)
\, \Qop_{\pm}^{-1}(\specbaz^2_{\pm};\twist) \, e^{-\vec \gamma\,  \vec  \SpinQ_{\text{tot}}}=
\, \Qop_{\pm}(\specbaz^1_{\pm};\twist) \, \Qop_{\pm}^{-1}(\specbaz^2_{\pm};\twist) \,\left( \Identity + \mathcal O_{\pm} (\twist)\right)\,.
\end{equation}
 The proof is trivial using that $\Qop_{\pm}(\specbaz^1,\twist) $ and 
 $\Qop_{\pm}^{-1}(\specbaz^2,\twist)$ commute.
 So the $\alg{su}(2)$ invariance in the $\twist \rightarrow 0$ limit 
 follows.

\section{More on \texorpdfstring{$\alg{sl}(2)$}{} transformation
  properties of Wronskian} 
\label{app:moresl2wronskian}

Let us investigate how  $\Qop_{\specsubtraction}(\specbaz;\twist)$ and $\Pop_{\specsubtraction}(\specbaz;\twist)$ 
transform under rotations of $\Qop_\pm$. One has
\begin{equation}
\begin{pmatrix}
\Qop_+ \\
\Qop_-
\end{pmatrix} \rightarrow
\begin{pmatrix}
\alpha \,& \beta\, \\
\gamma \,& \delta\,
\end{pmatrix}
\begin{pmatrix}
\Qop_+ \\
\Qop_-
\end{pmatrix}\, ,
\qquad {\rm with} \qquad 
\alpha\, \delta-\beta\, \gamma=1 \,.
\end{equation}
One then finds
\begin{equation}
\Pop_{\specsubtraction}(\specbaz;\twist) \rightarrow
\Pop_{\specsubtraction}(\specbaz;\twist)\, , \qquad
\Qop_{\specsubtraction}(\specbaz;\twist)  \rightarrow \Qop_{\specsubtraction}(\specbaz;\twist) + \Pop_{\specsubtraction}(\specbaz;\twist)\, \mathcal{O}^{\left(\begin{smallmatrix}
 \alpha & \beta\\
\gamma & \delta 
\end{smallmatrix}\right)}(\specsubtraction;\twist) \,.
\end{equation}
The explicit form of $\mathcal O$ is not important here, just notice that $\mathcal O$ vanishes for
\begin{equation}
\label{opersl2vanish}
\begin{pmatrix}
\alpha \,& \beta\, \\
\gamma \,& \delta\,
\end{pmatrix}= 
\begin{pmatrix}
\alpha \,& 0\, \\
0 \,& \alpha^{-1}\,
\end{pmatrix}\, ,
 \,\qquad
\begin{pmatrix}
\alpha \,& \beta\, \\
\gamma \,& \delta\,
\end{pmatrix}= 
\begin{pmatrix}
0 \,& i \, \alpha^{-1}\, \\
 i \, \alpha \,& 0\, 
\end{pmatrix}= \begin{pmatrix}
0 \,& i \, \\
 i  \,& 0\, 
\end{pmatrix}\, \begin{pmatrix}
\alpha \,& 0\, \\
0 \,& \alpha^{-1}\,
\end{pmatrix},
\end{equation}
which is the reason why equations \eqref{Ptauz}, \eqref{Qtauz} cannot be inverted for $\Qop_\pm$.
This freedom corresponds to the normalization of $\Qop_\pm$ and to the discrete operation of exchanging $\Qop_+$
with $\Qop_-$. 


\section{Numerics}
\label{app:numerics}
In this appendix we want to present more details on the method used in section \ref{sec:numerics} for finding the two-cut root distributions for both the $Q$ and the $P$ polynomials. We will perform it in two steps. Firstly, we will find all roots of the polynomial $Q(u)$ using the method presented in \cite{Bargheer:2008kj}. Then, using the information about the polynomial $Q(u)$ we will find $P(u)$ from the generalized Wronskian relation (\ref{TopjPQfinal}). We will end this appendix with a few figures presenting different aspects of the problem.

According to \cite{Bargheer:2008kj}, in order to find roots of the polynomial $Q(u)$ for given $M$ and $L\sim M$, we have to take as a starting point a slightly modified system. Let us fix the number of excitations $M$, and choose the length of the spin chain $L'\gg L$ such that $\frac{M}{L'} \ll \frac{1}{2}$. One finds that the approximate root distribution in this case is given in terms of the roots $z_k$ of Hermite polynomials as 
\begin{equation}\label{app:Bethe.roots}
u_k =\frac{1}{2\pi n_k }\left(L + iz_k \sqrt{2L} + \mathcal{O}( 1)\right) .
\end{equation}
Here, $z_k$ are the solutions of $H(z_k)=0$, where by $H$ we denoted a Hermite polynomial. To simplify our considerations we will investigate only the simplest two-cut solution with mode numbers given by $n_k=1$, for $k=1,\ldots,\frac{M}{2}$,  and $n_k=-1$, for $k=\frac{M}{2}+1,\ldots,M$. Knowing the approximate solution, we can use it as a starting point for the Newton method. This way we can find the solution of (\ref{betheeqsnotwist}) with desired precision for any $M$ and much larger $L'$. Unfortunately, this method does not work for the configurations close to half-filling because then the approximation (\ref{app:Bethe.roots}) is not good enough, and the Newton algorithm no longer converges. However, to overcome this difficulty, 
we can treat the root distribution as a function of the length, while keeping the magnon number $M$ fixed. Assuming in addition that the root distributions for slightly differing $L$'s should be numerically close, we can use the solution found for a given length as initial data for Bethe equations with smaller length. This way we get a sequence of configurations with fixed $M$ and decreasing $L$, corresponding to increasing filling $\frac{M}{L}$. This procedure is sufficient to produce the solutions of Bethe equations with excitation numbers very close to half-filling. In particular, we can get the configuration presented in figure \ref{L60M22} in the main text.

\begin{figure}
 \begin{center}
 \includegraphics[scale=0.92]{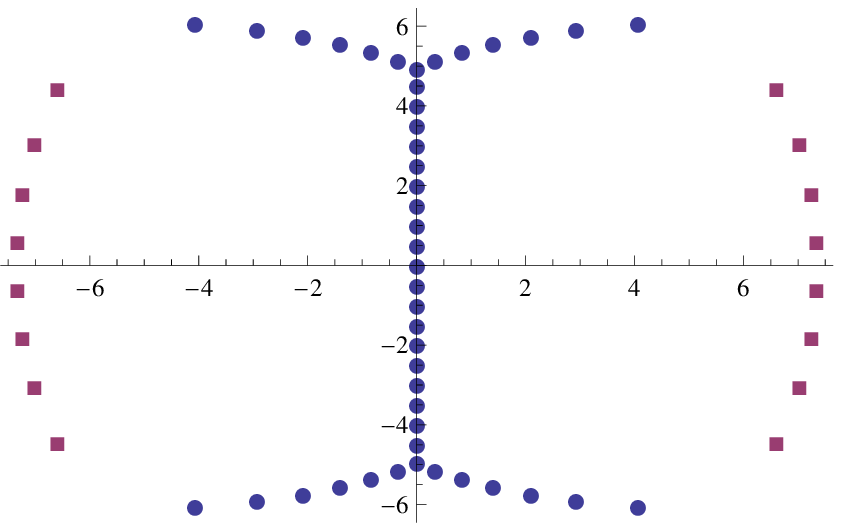} 

 (a)
 
 \bigskip
 \includegraphics[scale=0.92]{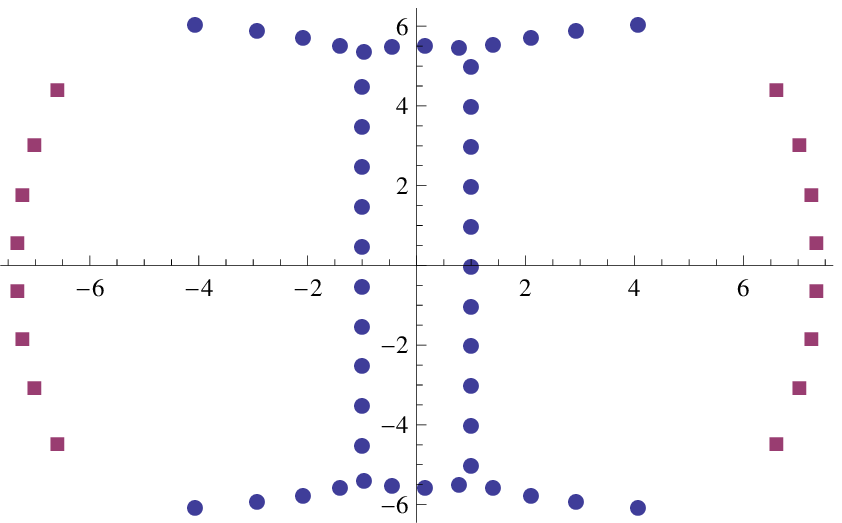}

(b) 

 \bigskip
 \includegraphics[scale=0.92]{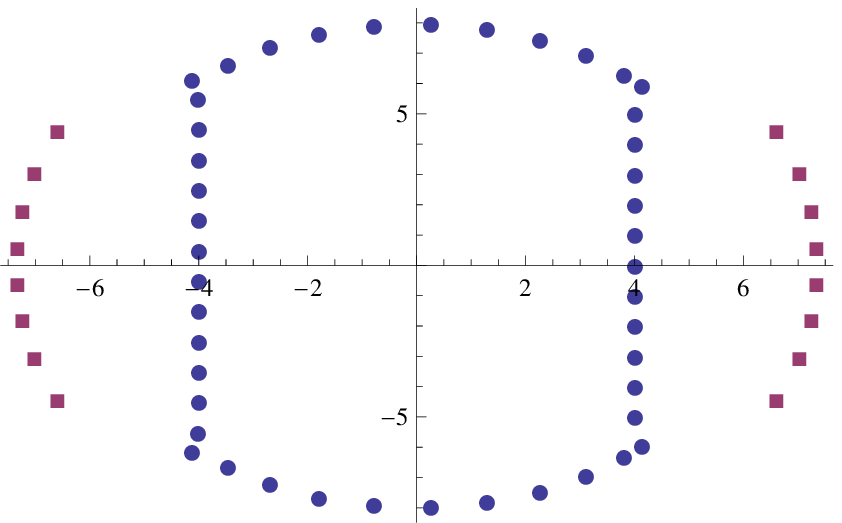}
 
 (c)
\end{center}
\caption{Root distribution of $Q(u)$ (purple squares) and $P(u)$ (blue dots) with various $\tilde{u}_0$ for $L=60$ and $M=16$.  (a) $\tilde{u}_0=0$, (b)~$\tilde{u}_0=1$, (c) $\tilde{u}_0=4$. Reading this figure from the bottom one can see that when $|\tilde{u}_0|$ is decreased the two condensates shown in (c) start to approach each other (b) and eventually assemble on one line (a). It is interesting to note that for even length $L$ the two condensates pass through each other without ``touching" (like in the figure), while for odd $L$ they ``scatter'', thereby producing double roots on the imaginary axis for $\tilde{u}_0=0$.}
\label{fig:various.u0}
\end{figure}

In order to get the root configuration of the P-polynomials we will use \eqref{TopjPQfinal} with $j=0$ and in the $\phi\to 0$ limit. At the left hand side we then have only $u^L$, while at the right hand side we can substitute $Q(u)=\prod_{i=1}^{\frac{M}{2}}(u^2-u_i^2)$. Here we denoted by $u_i$ all roots with positive real parts found at the previous stage. We assume the polynomial $P(u)$ to be of the form $P(u)=(u-\tilde{u}_0)\sum_{i=0}^{L-M}a_i\, u^i$
%
%
where we factorized out $(u-\tilde{u}_0)$ in order to use the freedom of $P(u)$ which we discussed in the main text. We can indeed always construct a solution of the untwisted Wronskian in the form $P'(u)=P(u)+\alpha\, Q(u)$
%
%
by choosing $\alpha$ such that any given $\tilde{u}_0$ become a root of $P'$. This way we can find a one-parameter family of solutions numerated by $\tilde{u}_0$. Now for given $\tilde{u}_0$ we substitute  the polynomial $P(u)$ of the mentioned form 
into the Wronskian relation. This way we will end up with the set of $L-M+1$ linear equations, where the coefficients $a_i$, $i=0,\ldots,L-M$ will play the role of unknowns. The solution of this linear equations is easy to find. It leaves us only with the problem of finding the roots of $P(u)$, which we can perform numerically. 
Using the methods presented above we are able to find configurations for chosen values of $L$ and $M$ below half-filling. Additionally, we may observe
%
the behaviour of the solution for different $\tilde{u}_0$, see also \cite{Gromov:2007ky}. In the main text we presented only the $\tilde{u}_0=0$ case. In figure \ref{fig:various.u0} we illustrate the root configurations also for other values.   


\end{document}